\documentclass[sigconf]{acmart}

\usepackage{indentfirst}
\usepackage{color}
\usepackage{enumitem}
\usepackage{multirow}
\usepackage{colortbl}
\usepackage{tabularray}
\usepackage{graphicx}
\usepackage{subfigure}
\usepackage{accents}
\usepackage{graphicx, nicefrac}
\usepackage{url}
\usepackage{bbm}
\usepackage{mdframed}
\usepackage[capitalize]{cleveref}

\usepackage{breakurl}
\usepackage[flushleft]{threeparttable}
\usepackage{bigstrut}

\usepackage[commandnameprefix=ifneeded]{changes}
\makeatletter
\AddToHook{cmd/added/before}{\def\Changes@AuthorColor{blue}}
\AddToHook{cmd/deleted/before}{\def\Changes@AuthorColor{red}}
\makeatother
\definechangesauthor[color=red]{removed}
\definechangesauthor[color=blue]{added}

\renewcommand{\added}[1]{{#1}}

\newtheorem{theorem}{Theorem}
\newtheorem{lemma}{Lemma}[section]

\newtheorem{remark}{Remark}
\newtheorem{assumption}{Assumption}
\usepackage[ruled,linesnumbered,vlined]{algorithm2e}

\usepackage{amsmath,amsfonts,bm}

\def\eqref#1{equation~\ref{#1}}
\def\Eqref#1{Eq.(\ref{#1})}

\def\1{\bm{1}}

\def\indicator#1{{{\mathbbm{1}}}_{{#1}}}
\def\betafunc#1{{{\mathrm{B}}}\left({{#1}}\right)}

\def\ve{{\bm{e}}}

\def\vx{{\mathbf{x}}}
\def\hvx{{\hat\vx}}
\def\hx{{\hat x}}
\def\vy{{\mathbf{y}}}
\def\hvy{{\hat\vy}}
\def\hy{{\hat y}}

\def\nny{{\overline{n}_y}}
\def\nnx{{\overline{n}_x}}

\def\mE{{\mathbf{E}}}

\DeclareMathAlphabet{\mathsfit}{\encodingdefault}{\sfdefault}{m}{sl}
\SetMathAlphabet{\mathsfit}{bold}{\encodingdefault}{\sfdefault}{bx}{n}

\newcommand{\eg}{{\it e.g.}}
\newcommand{\etal}{{\it et al.}}
\newcommand{\ie}{{\it i.e.}}
\newcommand{\stitle}[1]{\vspace{2mm} \noindent {\bf #1}}
\newcommand\method[1]{\textsc{#1}}
\newcommand\ours{$\mathcal{B}^4$}

\AtBeginDocument{%
  }

\copyrightyear{2024}
\acmYear{2024}
\setcopyright{acmlicensed}\acmConference[ASE '24]{39th IEEE/ACM
International Conference on Automated Software Engineering }{October
27-November 1, 2024}{Sacramento, CA, USA}
\acmBooktitle{39th IEEE/ACM International Conference on Automated Software
Engineering (ASE '24), October 27-November 1, 2024, Sacramento, CA, USA}
\acmDOI{10.1145/3691620.3695536}
\acmISBN{979-8-4007-1248-7/24/10}

\begin{document}

\title{\ours: Towards Optimal Assessment of Plausible Code Solutions with Plausible Tests}

\author{Mouxiang Chen}
\affiliation{
  \institution{The State Key Laboratory of Blockchain and Data Security, Zhejiang University}
  \city{Hangzhou}
  \country{China}}
\email{chenmx@zju.edu.cn}

\author{Zhongxin Liu}
\affiliation{
  \institution{The State Key Laboratory of Blockchain and Data Security, Zhejiang University}
  \city{Hangzhou}
  \country{China}}
  \authornote{Corresponding Author}
  \authornote{Also with Hangzhou High-Tech Zone (Binjiang) Institute of Blockchain and Data
Security}
\email{liu_zx@zju.edu.cn}

\author{He Tao}
\affiliation{
  \institution{Zhejiang University}
  \city{Hangzhou}
  \country{China}}
\email{tao_he@zju.edu.cn}

\author{Yusu Hong}
\affiliation{
  \institution{Zhejiang University}
  \city{Hangzhou}
  \country{China}}
\email{yusuhong@zju.edu.cn}

\author{David Lo}
\affiliation{
  \institution{Singapore Management University}
  \city{Singapore}
  \country{Singapore}}
\email{davidlo@smu.edu.sg}

\author{Xin Xia}
\affiliation{
  \institution{Zhejiang University}
  \city{Hangzhou}
  \country{China}}
\email{xin.xia@acm.org}

\author{Jianling Sun}
\affiliation{
  \institution{The State Key Laboratory of Blockchain and Data Security, Zhejiang University}
  \city{Hangzhou}
  \country{China}}
\email{sunjl@zju.edu.cn}

\begin{abstract}
  Selecting the best code solution from multiple generated ones is an essential task in code generation, which can be achieved by using some reliable validators (\eg, developer-written test cases) for assistance. Since reliable test cases are not always available and can be expensive to build in practice, researchers propose to automatically generate test cases to assess code solutions. However, when both code solutions and test cases are plausible and not reliable, selecting the best solution becomes challenging. Although some heuristic strategies have been proposed to tackle this problem, they lack a strong theoretical guarantee and it is still an open question whether an optimal selection strategy exists. Our work contributes in two ways. First, we show that within a Bayesian framework, the optimal selection strategy can be defined based on the posterior probability of the observed passing states between solutions and tests. The problem of identifying the best solution is then framed as an integer programming problem. Second, we propose an efficient approach for approximating this optimal (yet uncomputable) strategy, where the approximation error is bounded by the correctness of prior knowledge. We then incorporate effective prior knowledge to tailor code generation tasks. Both theoretical and empirical studies confirm that existing heuristics are limited in selecting the best solutions with plausible test cases. Our proposed approximated optimal strategy \ours~significantly surpasses existing heuristics in selecting code solutions generated by large language models (LLMs) with LLM-generated tests, achieving a relative performance improvement by up to 50\% over the strongest heuristic and 246\% over the random selection in the most challenging scenarios. Our code is publicly available at \url{https://github.com/ZJU-CTAG/B4}.
\end{abstract}




\maketitle

\section{Introduction}

\begin{figure}[t]
    \centering
    \includegraphics[width=0.45\textwidth]{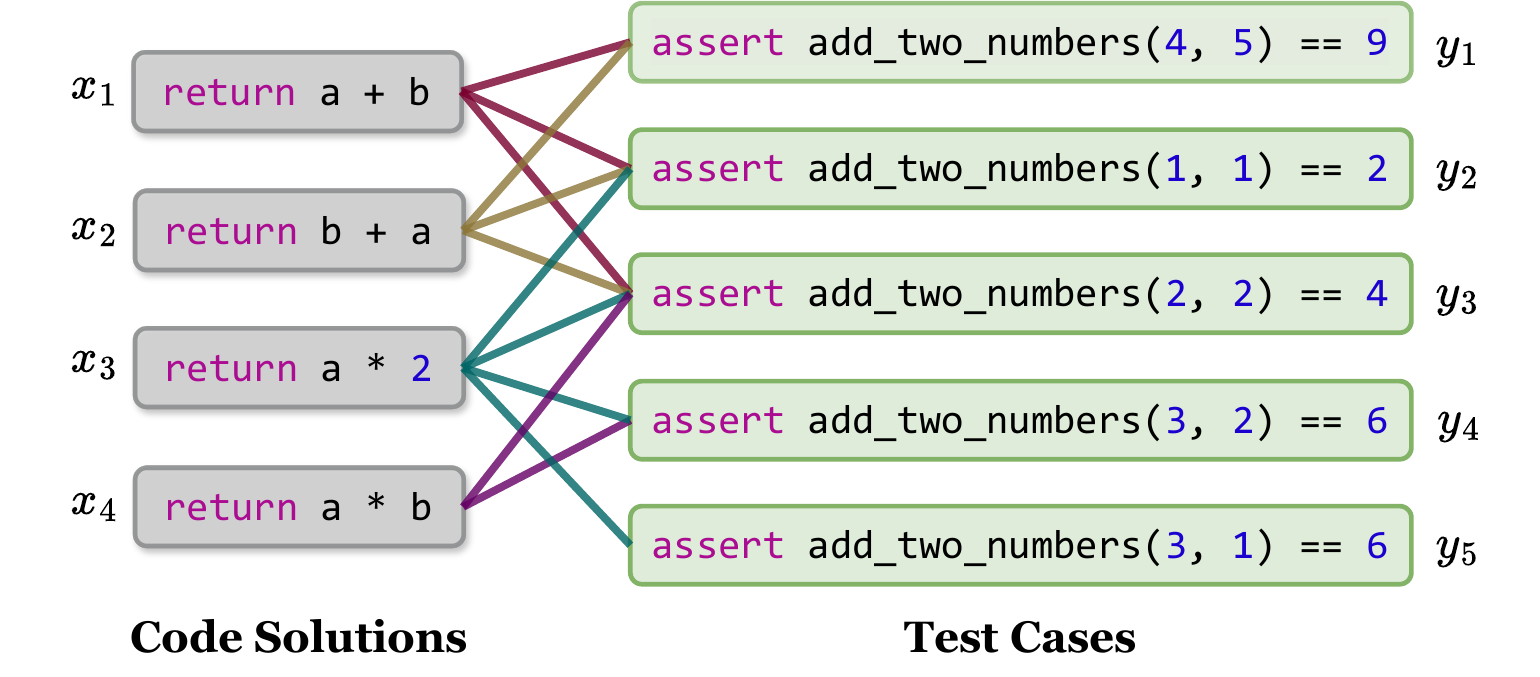}
    \vspace{-2mm}
    \caption{A simple example showing the problem "return the sum of $a$ and $b$". A link between a generated code solution and a generated test case indicates that the solution passes the test. How can we select the best code solution solely based on these links?}
    \label{fig:example}
    \vspace{-5mm}
\end{figure}

Code generation is an important task in the field of software engineering \cite{liu2022deep}, aiming to generate code solutions that satisfy the given requirement. In practice, we often face the problem of selecting the best code solution from multiple generated alternatives \cite{chen2021evaluating,li2022competition}. A common practice is using some validators (\eg, test cases) to assess the validity of each solution and choose the best one \cite{chen2023codet,yang2017better,roziere2022leveraging,shi2022natural}. However, in real-world scenarios, reliable test cases are not always available. Developing and maintaining reliable test cases can also be resource-intensive and laborious. With advancements in deep learning and large language models (LLMs), using auto-generated test cases has gained popularity among researchers and practitioners \cite{9401982, 10172800, 10172590, 10329992}. Unfortunately, selecting code solutions based on these potentially unreliable tests poses significant challenges, since incorrect test cases can disrupt our decision-making. \cref{fig:example} provides an example, where selecting the best code solution becomes difficult since the fourth and fifth test cases are incorrect.

Few studies systematically explore how to assess \textit{plausible} code solutions and select the best using \textit{plausible} test cases. Under the assumption that the generated test cases are (mostly) correct, some existing research favors the solutions that pass the most test cases \cite{lahiri2023interactive,li2022competition,le2022coderl,roziere2022leveraging}. However, this strategy is ineffective when test cases are merely plausible, indicated by our theoretical analysis (see \cref{sec:theorem}). Other research addresses this challenge by designing clustering-based heuristic rules. For instance, Shi \etal~\cite{shi2022natural} and Li \etal~\cite{li2022competition} clustered code solutions based on test outputs, and selected the solutions from the largest cluster. Chen \etal~\cite{chen2023codet} similarly clustered code solutions based on the passed test cases, and selected the best cluster according to the count of solutions and passed test cases in each. However, these heuristics rely on human-designed rules and lack strong theoretical foundations, leading to potentially suboptimal performance. To the best of our knowledge, the optimal selection strategy for this problem is still an open question.

In this work, we aim to develop a general framework to define and compute the optimal selection strategy. We first show that under a Bayesian framework, the optimal strategy can be defined based on the posterior probability of the observed passing states between solutions and tests. The problem of identifying the optimal strategy is then framed as an integer programming problem. Under a few assumptions, this posterior probability can be further expanded into four integrals, which cannot be directly computed due to four unknown prior distributions. We then leverage Bayesian statistics techniques to deduce a computable form for approximating this posterior probability and optimize the integer programming from exponential to polynomial complexity. The approximation error is bounded by the correctness of prior knowledge. Based on this bound, we investigate two effective priors and incorporate them into our framework to enhance code generation performance. Given that the approximated optimal strategy involves scoring code solutions with four Beta functions \cite{davis1972gamma}, we refer to it as \ours.

Based on our developed framework, we further provide a theoretical analysis to compare \ours~with existing heuristics. We observe that some heuristics require sufficient correct test cases, while others necessitate a higher probability of correct code solutions, as confirmed by subsequent simulated experiments. In real-world applications involving selecting LLM-generated code solutions with LLM-generated test cases, \ours~significantly outperforms existing heuristics across five LLMs and three benchmarks.

In summary, our paper makes the following contributions:
\begin{itemize}[leftmargin=*]
    \item \textbf{Optimal Strategy.} We systematically address the challenging problem of selecting plausible code solutions with plausible tests and establish an optimal yet uncomputable strategy.
    \item \textbf{Technique.} We derive an efficiently computable approach to approximate the uncomputable optimal strategy with an error bound. While our framework is broadly applicable, we adapt it to code generation by incorporating two effective priors.
    \item \textbf{Theoretical Study.} Using our framework, we explore the conditions under which existing heuristics are effective or ineffective and compare them to the approximated optimal strategy \ours.
    \item \textbf{Empirical Study.} We empirically evaluate our selection strategy with five code LLMs on three benchmarks. Experimental results show that our strategy demonstrates up to a 12\% average relative improvement over the strongest heuristic and a 50\% improvement in the most challenging situations where there are few correct solutions.
\end{itemize}


\section{Preliminaries}

\paragraph{Notations}

We use bold lowercase
letters to denote vectors (\eg, $\vx$ and $\vy$), bold uppercase letters to denote matrices (\eg, $\mE$), and thin letters to denote scalars (\eg, $x$ and $y$). We also use thin uppercase letters to denote random variables (\eg, $X$, $Y$, and $E$). $\ve_i$ denotes the $i$-th row in matrix $\mE$. The index set $[N]$ denotes $\{1,2,\cdots,N\}$. $\{0,1\}^N$ denotes a length-$N$ binary vector, and $\{0,1\}^{N\times M}$ denotes an $N\times M$ binary matrix.

\subsection{Problem Definition}

Code generation is a crucial task in software engineering, which aims at automatically generating a code solution $x$ from a given context $c$. We explore the selection of the best code solution from $N$ code solutions generated based on $c$, with $M$ test cases (also generated based on $c$) to aid this selection. It is worth noting that the correctness of both code solutions and test cases is \textit{plausible}; they might be either correct or incorrect, which is unobserved however. All we can observe is a matrix $\mE=\{e_{ij}\}_{N\times M} \in\{0,1\}^{N\times M}$ where $e_{ij}=1$ indicates the $i$-th code solution passes the $j$-th test case, and 0 indicates failure. We term $\mE$ as \textbf{passing matrix}, and $e_{ij}$ as \textbf{passing state}.

Let $\vx=\{x_1,\cdots, x_N\}\in \{0,1\}^{N}$ denote the ground-truth correctness of each code solution (unknown to us), in which 1 denotes \textit{correct} and 0 denotes \textit{incorrect}. We assume at least one code solution is correct since designing a selection strategy would be meaningless without any correct code. Similarly, the correctness of each test case is denoted by $\vy=\{y_1,...,y_M\}\in \{0,1\}^{M}$. We assume that all correct code solutions share identical \textit{functionality} and all tests are not flaky, meaning that all solutions pass the same test cases on the same context $c$. This can be formulated as the following assumption.

\begin{assumption}[Consistency]\label{asm:identical_functionality} For all $i,j\in[N]$, if $x_i=1$ and $x_j=1$, then $\mE$ should satisfy:
    \begin{align*}
    \ve_i=\ve_j\quad (\ie, e_{ik}=e_{jk},\quad \forall k\in [M]).
    \end{align*}
    Furthermore, the correctness of test cases $\vy$ corresponds to the passing states of the correct code solutions. Formally, if $x_i=1, i\in [N]$, then:
    \begin{align*}
    \vy = \ve_i\quad (\ie, y_k=e_{ik},\quad \forall k\in [M]).
    \end{align*}
\end{assumption}

This assumption indicates that $\mE$ and $\vy$ should be \textit{consistent} with $\vx$. Intuitively, $\mE$ should satisfy that the rows corresponding to the correct code solutions are the same. $\vy$ is defined based on these rows. For example, in \cref{fig:example}, we have $\vx=\{1,1,0,0\}$, $\vy=\{1,1,1,0,0\}$, and
\begin{align}
    \label{eq:passing_matrix_example}
    \mE=\begin{pmatrix}
    1 & 1 & 1 & 0 & 0 \\
    1 & 1 & 1 & 0 & 0 \\
    0 & 1 & 1 & 1 & 1 \\
    0 & 0 & 1 & 1 & 0
\end{pmatrix}.
\end{align}

In this paper, our goal is to use $\mE$ to assess the correctness of code solutions and select the best one by recovering $\vx$ and $\vy$ from $\mE$. Following Chen \etal~\cite{chen2023codet}, we do not rely on any specific details of the code solutions or test cases in this paper.

\subsection{Existing Heuristics}

In this section, we briefly review two representative heuristic methods for addressing this problem. The first family of methods \method{MaxPass} \cite{lahiri2023interactive,li2022competition,le2022coderl,roziere2022leveraging} always rewards passing test cases. The best code solution can be selected by counting the passed cases, \ie, 
\begin{align*}
    \text{Select code solution }i, \text{ where }i = \mathop{\arg\max}_{i\in[N]} \sum_{j=1}^M e_{ij}.
\end{align*}

The other family of methods examines the consensus between code solutions and test cases, and clusters the code solutions with the same functionality \cite{li2022competition,shi2022natural,chen2023codet}. One of the most representative methods is \method{CodeT} \cite{chen2023codet}. It divides the code solutions into $K$ disjoint subsets based on functionality: $S^x=\{S_1^x,\cdots, S_K^x\}$, where each set $S^x_i$ ($i\in[K]$) consists of code solutions that pass the same set of test cases, denoted by $S_i^y$. The tuple ($S_i^x$, $S_i^y$) is termed a \textit{consensus set}. Taking \cref{fig:example} as an example, there are three consensus sets: $(\{x_1,x_2\}, \{y_1,y_2,y_3\})$, $(\{x_3\}, \{y_2,y_3,y_4,y_5\})$ and $(\{x_4\}, \{y_3,y_4\})$. 

\method{CodeT} proposes that a consensus set containing more code solutions and test cases indicates a higher level of consensus, and thus the more likely they are correct. Therefore, \method{CodeT} scores each consensus set based on the capacity and selects the code solutions associated with the highest-scoring set, \ie,
\begin{align*}
    \text{Select code solutions }i \in S_k, \text{ where }k =\mathop{\arg\max}_{k\in[K]} |S_k^x|\cdot|S_k^y|.
\end{align*}

Similarly, other clustering methods, such as MBR-\method{exec} \cite{shi2022natural} and AlphaCode-C \cite{li2022competition}, also cluster the code solutions based on test cases, but only score each set by the number of code solutions $|S_k^x|$. We focus our analysis on \method{CodeT} as it was verified to outperform other existing scoring strategies \cite{chen2023codet}.


In this study, we develop a systematic analysis framework, to evaluate the effectiveness of these heuristics and address the following research questions (RQs):

\begin{itemize}[leftmargin=*]
    \item \textit{RQ1: Given a passing matrix $\mE$, what constitutes the optimal selection strategy?}
    \item \textit{RQ2: Is this optimal strategy computable?}
    \item \textit{RQ3: Can a practical algorithm be developed to compute (or approximate) this optimal strategy efficiently?}
    \item \textit{RQ4: Under what conditions do existing heuristics not work, based on our developed analysis framework?}
    \item \textit{RQ5: If the answer to RQ3 is true, how does the computable (or approximated) optimal strategy compare to these heuristics?}
\end{itemize}
\section{Methodology}

In this section, we outline our proposed methodology to address this problem. 

\subsection{Optimal Strategy}\label{sec:map}

We use $X=\{X_1,\cdots,X_N\}\in \{0,1\}^N$, $Y=\{Y_1,\cdots,Y_M\}\in \{0,1\}^M$, and $E=\{E_{ij}\}_{N\times M}\in \{0,1\}^{N\times M}$ to denote random variables of code solutions' and tests' correctness, and the passing matrix, respectively. Note that all $X$, $Y$, and $E$ depend on the same context $C$, which we omit for ease of notation. A strategy's estimation for $X$ and $Y$ is denoted by $\hvx=\{\hx_1,\cdots,\hx_N\}$ and $\hvy=\{\hy_1,\cdots,\hy_M\}$. To answer \textit{RQ1}, our goal is to find \textit{the most probable} $\hvx$ and $\hvy$ given an observation $E=\mE$. This motivates us to design the optimal strategy by modeling $P(X, Y\mid E)$. Based on Bayes' theorem, we have:
\begin{align*}
    \underbrace{P(X, Y\mid E)}_{\text{posterior}}=\frac{P(E\mid X, Y)}{P(E)}{P(X, Y)}\propto \underbrace{P(E\mid X, Y)}_{\text{likelihood}} \underbrace{P(X, Y)}_{\text{prior}}.
\end{align*}

Therefore, we propose to use \textbf{maximum a posteriori} (MAP) estimator to obtain the best solution \cite{degroot2005optimal}:
{\small
\begin{align}
    \label{eq:map}
    \hvx^*, \hvy^*=\mathop{\arg\max}_{\hvx\in\{0,1\}^N, \hvy\in\{0,1\}^M} \underbrace{P(E=\mE\mid X=\hvx, Y=\hvy)}_{\text{likelihood}} \underbrace{P(X=\hvx, Y=\hvy)}_{\text{prior}}.
\end{align}
}

That is to say, we exhaustively explore all $2^N$ possible configurations of $\hvx$ and $2^M$ configurations of $\hvy$, computing the likelihood and prior for each pair. We then find the $\hvx^*$ and $\hvy^*$ that yield the highest posterior and select the correct code solutions and test cases indicated by $\hvx^*$ and $\hvy^*$. This optimization problem is a 0/1 integer programming problem, in which all variables are restricted to 0 or 1. The following then answers \textit{RQ1}.

\vspace{1.5mm}
\begin{mdframed}[linecolor=black,linewidth=1pt]
\vspace{-0.5mm}
\textbf{Answer to \textit{RQ1}: } Given a passing matrix $\mE$, the optimal selection strategy can be framed as a 0/1 integer programming problem, by finding the one $\hvx\in\{0,1\}^N$ and $\hvy\in\{0,1\}^M$ that maximizes the posterior probability $P(X=\hvx, Y=\hvy \mid E=\mE)$.
\vspace{-3.5mm}
\end{mdframed}

\vspace{1.5mm}

Before calculating \Eqref{eq:map}, we first introduce the following two assumptions which are necessary for our subsequent computation. 

\begin{assumption}
    \label{asm:independence_xy}
    The code solutions $X$ and the test cases $Y$ are independent and randomly sampled.
\end{assumption}

\begin{assumption}
    \label{asm:dependence_e}
    Each $E_{ij}$ is only dependent by the $X_i$ and $Y_j$, $\forall i\in [N], j\in[M]$.
\end{assumption}
\begin{remark}\label{rem:assumption}
\cref{asm:independence_xy} is also used by Chen \etal~\cite{chen2023codet}. \cref{asm:dependence_e} assumes that a passing state $E_{ij}$ is independent of any other variables except for the corresponding code $X_i$ and test case $Y_j$, \added{which means that $E_{ij} (i \in [N], j \in [M])$ are conditional independent when given $X$ and $Y$}. We will further discuss these assumptions in \cref{sec:discussion}.
\end{remark}
Based on \cref{asm:dependence_e}, we can explicitly formulate $P(E_{ij}\mid X_i, Y_j)$ as follows,
{\small
\begin{equation}
\begin{aligned}
    P(E_{ij}=1\mid X_i=1, Y_j=1)=1,&\quad
    P(E_{ij}=1\mid X_i=1, Y_j=0)=0,\\
    \label{eq:eij}
    P(E_{ij}=1\mid X_i=0, Y_j=1)=\theta_1,&\quad
    P(E_{ij}=1\mid X_i=0, Y_j=0)=\theta_0,
\end{aligned}
\end{equation}
}%
where $\theta_1$ and $\theta_0$ are unknown parameters, indicating the probabilities of an incorrect solution passing a correct test case ($\theta_1$) and passing an incorrect test case ($\theta_0$). \Eqref{eq:eij} suggests that if a solution is correct ($X_i=1$), $E_{ij}$ is \textit{deterministic} by $Y_j$ to fulfill the consistency (\cref{asm:identical_functionality}). When a solution is incorrect ($X_i=0$), $E_{ij}$ is a Bernoulli \textit{random} variable, \ie, a random variable that can only take 0 or 1, where the probability depends on $Y_j$.

\begin{figure}[t]
    \centering
    \includegraphics[width=0.45\textwidth]{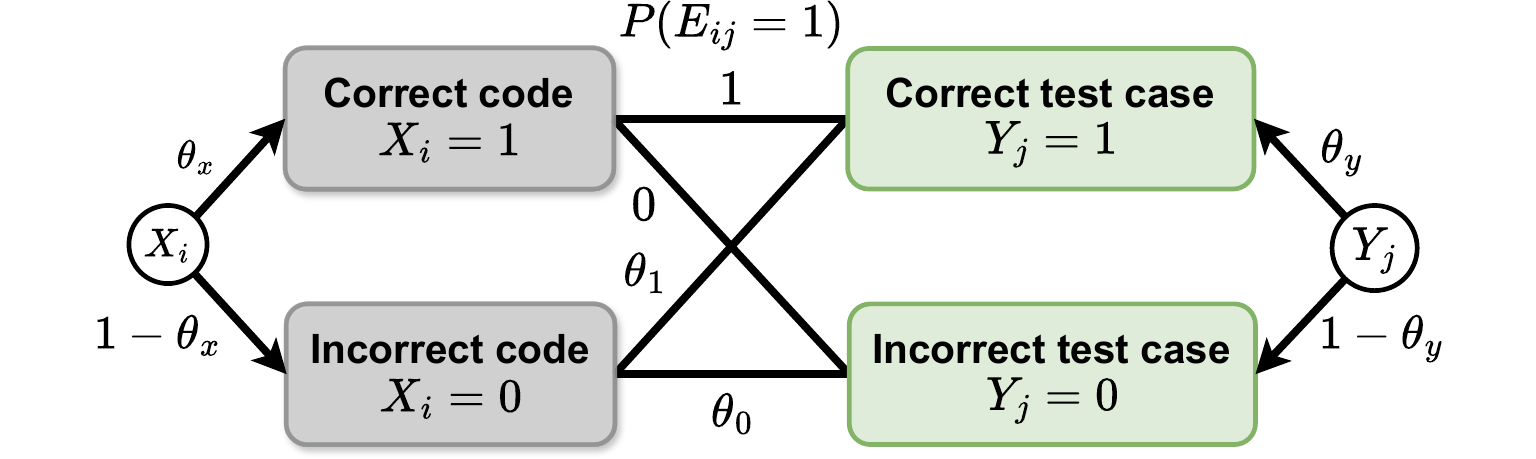}
    \vspace{-2mm}
    \caption{Illustration of the generation process. The correctness of code $X_i$ and test case $Y_j$ is sampled using parameters $\theta_x$ and $\theta_y$ respectively. $E_{ij}$ is generated based on $X_i$ and $Y_j$, using the corresponding parameters (1, 0, $\theta_1$ or $\theta_0$).}
    \label{fig:intro-method}
    \vspace{-5mm}
\end{figure}

Based on Assumption~\ref{asm:independence_xy}, the correctness of code solutions $X$ and test cases $Y$ are independent and therefore follow Bernoulli distributions as well. Suppose that:
\begin{align*}
    P(X_i=1)=\theta_x,\quad P(Y_j=1)=\theta_y,\quad\forall i\in[N],j\in[M],
\end{align*}
where $\theta_x$ and $\theta_y$ are two unknown parameters. To summarize, \cref{fig:intro-method} illustrates the generation process of $E$ based on four unknown parameters $\theta_1$, $\theta_0$, $\theta_x$ and $\theta_y$ for a clear presentation. 

For ease of notation, we omit the random variables in the probability expressions in subsequent sections, \eg, using $P(\hvx, \hvy)$ to replace $P(X=\hvx, Y=\hvy)$. In the following sections, we provide a detailed explanation of how to derive the likelihood and prior in \Eqref{eq:map} based on the generation process proposed in \cref{fig:intro-method}.

\stitle{Computing the likelihood.}
Based on \cref{asm:dependence_e} and \cref{rem:assumption}, we can expand the likelihood $P(\mE\mid \hvx, \hvy)$ into the following form:
\begin{align}
    \nonumber
    P(\mE\mid \hvx, \hvy)&=\prod_{i}\prod_{j} 
    P(e_{ij}\mid \hvx, \hvy)\\
    \label{eq:likelihood}
    &=\underbrace{
        \prod_{\hx_i=1}\prod_{j} 
     P(e_{ij}\mid \hvx, \hvy)}_{P_1}
    \underbrace{
        \prod_{\hx_i=0}\prod_{j} 
     P(e_{ij}\mid \hvx, \hvy)
    }_{P_0},
    \end{align}
where $i\in[N]$ and $j\in[M]$. The first equality is based on the independence of $E_{ij}$. The second equality splits $e_{ij}$ into two parts, \ie, $P_1$ and $P_0$, based on $\hx_i$.

According to \Eqref{eq:eij}, $P_1$ is either 1 or 0. If $\hvy$ and $\mE$ are consistent with $\hvx$ (\ie, satisfy \cref{asm:identical_functionality}), then $P_1$ is 1; otherwise $P_1$ is 0. Here we only focus on consistent configurations that satisfy \cref{asm:identical_functionality}. Under this condition, $P(\mE\mid \hvx, \hvy) = P_0$, so we only need to compute $P_0$. Suppose:
\begin{equation}
\begin{aligned}
    \mE_{1}=\{e_{ij}\mid \hx_i=0, \hy_j=1, i\in[N], j\in[M]\},\\
    \label{eq:e12}
    \mE_{0}=\{e_{ij}\mid \hx_i=0, \hy_j=0, i\in[N], j\in[M]\}.
\end{aligned}
\end{equation}

Based on \Eqref{eq:eij}, $\mE_{1}$ (or $\mE_0$) contains a set of independent Bernoulli variables related to $\theta_1$ (or $\theta_0$). Therefore: 
{\small
\begin{align}
    \nonumber
    P_0&=
        \prod_{\hx_i=0}\prod_{\hy_j=1}
    P(e_{ij}\mid \hvx, \hvy)
        \cdot \prod_{\hx_i=0}\prod_{\hy_j=0} 
    P(e_{ij}\mid \hvx, \hvy)\\
    \nonumber
    &=P(\mE_{1}\mid \hvx, \hvy)\cdot P(\mE_{0}\mid \hvx, \hvy)\\
    \nonumber
    &=\int_0^1P(\mE_{1} \mid \theta_1)P({\theta_1})\mathrm{d}\theta_1
    \int_0^1P(\mE_{0} \mid \theta_0)P({\theta_0})\mathrm{d}\theta_0\\
    \label{eq:p0}
    &=\int_0^1\theta_1^{n_1} (1-\theta_1)^{|\mE_1|-n_1} P({\theta_1})\mathrm{d}\theta_1
    \int_0^1\theta_0^{n_0} (1-\theta_0)^{|\mE_0|-n_0} P({\theta_0})\mathrm{d}\theta_0,
\end{align}
}%
where the third equality uses the fact that $\mE_1$ only depends on $\theta_1$ and $\mE_0$ only depends on $\theta_0$, which follows Bernoulli distributions based on \Eqref{eq:eij}. We leverage the law of total probability, where $P(\theta_1)$ and $P(\theta_0)$ are prior distributions for the two unknown parameters. The fourth equality leverages the formulation of the Bernoulli distribution, where $n_1=\sum_{e_{ij} \in \mE_1} e_{ij}$ and $n_0=\sum_{e_{ij} \in \mE_0} e_{ij}$ are the element sums of $\mE_1$ and $\mE_0$ respectively.


\stitle{Computing the prior.}
To compute the prior $P(\hvx,\hvy)$, following the similar derivation as above, we have:
{\small
\begin{align}
    \nonumber
    &P(\hvx, \hvy)=P(\hvx)P(\hvy)\\
    \nonumber
    &=\int_0^1P(\hvx \mid \theta_x)P({\theta_x})\mathrm{d}\theta_x
    \int_0^1P(\hvy \mid \theta_y)P({\theta_y})\mathrm{d}\theta_y\\
    \label{eq:prior_xy}
    &=\int_0^1\theta_x^{n_x} (1-\theta_x)^{N-n_x} P({\theta_x})\mathrm{d}\theta_x \int_0^1\theta_y^{n_y} (1-\theta_y)^{M-n_y} P({\theta_y})\mathrm{d}\theta_y,
\end{align}
}%
where $P(\theta_x)$ and $P(\theta_y)$ are prior distributions. $n_x=\sum_{\hx_i\in\hvx} \hx_i$ and $n_y=\sum_{\hy_j\in\hvy} \hy_j$ are the element sums of $\hvx$ and $\hvy$, respectively. 

\vspace{1.5mm}

\begin{mdframed}[linecolor=black,linewidth=1pt]
\vspace{-0.5mm}
\textbf{Answer to \textit{RQ2}: } Under Assumptions \ref{asm:independence_xy} and \ref{asm:dependence_e}, the posterior of the optimal strategy can be expanded into four integrals (\Eqref{eq:p0} and \Eqref{eq:prior_xy}) related to some \textit{observed} events ($n_1$, $n_0$, $n_x$, and $n_y$) and prior distributions on four \textit{unobserved} parameters ($\theta_1$, $\theta_0$, $\theta_x$, and $\theta_y$), which is not computable.
\vspace{-0.5mm}
\end{mdframed}

\subsection{Practical Implementation}\label{sec:efficient}

Recall that to compute the optimal strategy, we need to compute likelihood (\Eqref{eq:p0}) and prior (\Eqref{eq:prior_xy}), which is not computable however due to complicated integrals and unknown prior distributions. In this section, we describe how to design an efficient approach to approximate the optimal strategy.

\stitle{Computing integrals.}
In Bayesian statistics, employing \textit{conjugate distributions} for prior distributions is a standard technique to simplify integrals in posterior computation \cite{raiffa2000applied}. In our case, all the variables $X$, $Y$, and $E$ follow the Bernoulli distributions, whose conjugate prior is the Beta distribution \cite{bayes1763lii}. Thus, we assume the four parameters follow \textbf{Beta distributions}, formally,
\begin{equation}
\begin{aligned}
    P(\theta_0)\propto \theta_0^{\alpha_0-1}(1-\theta_0)^{\beta_0-1},&\quad P(\theta_1)\propto \theta_1^{\alpha_1-1}(1-\theta_1)^{\beta_1-1},\\
    \label{eq:theta_prior}
    P(\theta_x)\propto \theta_x^{\alpha_x-1}(1-\theta_x)^{\beta_x-1},&\quad P(\theta_y)\propto \theta_y^{\alpha_y-1}(1-\theta_y)^{\beta_y-1},
\end{aligned}
\end{equation}
where $\alpha$ and $\beta$ are eight hyperparameters that reflect our existing belief or prior knowledge. We ignore all probability normalizing constants for ease of notation since they will not change the selection decision. These hyperparameters allow us to integrate some effective prior knowledge, which will be elaborated in \cref{sec:prior}.

To illustrate how Beta distributions simplify computation, we take $\theta_x$ as an example. Combining the integral about $\theta_x$ in \Eqref{eq:prior_xy} with $P(\theta_x)$ in \Eqref{eq:theta_prior}, we obtain:
\begin{align*}
    &\int_0^1\theta_x^{n_x} (1-\theta_x)^{N-n_x} P({\theta_x})\mathrm{d}\theta_x \\
    \propto &\int_0^1\theta_x^{n_x} (1-\theta_x)^{N-n_x} \theta_x^{\alpha_x-1}(1-\theta_x)^{\beta_x-1} \mathrm{d}\theta_x \\
    = & \int_0^1\theta_x^{n_x+\alpha_x-1} (1-\theta_x)^{N-n_x + \beta_x - 1}\mathrm{d}\theta_x\\
    = &\ \betafunc{n_x+\alpha_x, N-n_x+\beta_x},
\end{align*}
where $\betafunc{\cdot}$ is known as the \textit{Beta function} \cite{davis1972gamma}, which can be efficiently computed by modern scientific libraries like SciPy \cite{pauli2020scipy}. This deduction is applicable to $\theta_1$, $\theta_0$, and $\theta_y$ as well. Combining \Eqref{eq:map}, \Eqref{eq:likelihood}, \Eqref{eq:p0}, and \Eqref{eq:prior_xy}, and applying the similar transformation to integrals yields the formula for the computable posterior:
\begin{align}
    \nonumber
    P&(\mE\mid \hvx, \hvy)P(\hvx, \hvy)=P_1\cdot P_0\cdot P(\hvx, \hvy)\\
    \nonumber
    \propto P_1\cdot &\left[
        \betafunc{n_1+\alpha_1, |\mE_1|-n_1+\beta_1} \betafunc{n_0+\alpha_0, |\mE_0|-n_0+\beta_0}
    \right]\\
    \label{eq:posterior_four_b_p1}
    \cdot &\left[
        \betafunc{n_x+\alpha_x, N-n_x+\beta_x} \betafunc{n_y+\alpha_y, M-n_y+\beta_y}
    \right]
\end{align}


This formula implies that the posterior probability can be approximated by multiplying four Beta functions, multiplied by a term $P_1$ indicating whether $\hvx$, $\hvy$, and $\mE$ are consistent. We next present an error bound for this approximation (Proof can be found in the online Appendix \cite{zenodo}).

\begin{theorem}[Approximation error bound]
    \label{thm:bound_b4}
    Let $\Delta$ denote the absolute error between the true posterior (\ie, $P(\hvx, \hvy\mid \mE)$) and the estimated posterior probability (\ie, multiplying the four Beta functions with the probability normalizing constants in \Eqref{eq:theta_prior}). Then:
    \begin{align*}
        \Delta\leq \frac{2}{P(\mE)}\left(c_1 \Delta_{\theta_1}+c_0 \Delta_{\theta_0}+c_x \Delta_{\theta_x}+c_y \Delta_{\theta_y}\right),
    \end{align*}
    where $\Delta_{\theta_1}$ is the total variance distance \cite{tsybakov2008introduction} between $P(\theta_1)$ and our assumed Beta prior distribution for $\theta_1$. $\Delta_{\theta_0}$, $\Delta_{\theta_x}$, and $\Delta_{\theta_y}$ are defined similarly. $c_1$, $c_0$, $c_x$, and $c_y$ are some positive constants less than 1.
\end{theorem}

\cref{thm:bound_b4} shows that the difference of scores given by the approximated approach and the optimal strategy (\ie, the true posterior probability) is bounded by the approximation errors in the prior distributions of the four parameters. If we can accurately give the prior distributions for each parameter $\theta$, then $\Delta_{\theta_1}=\Delta_{\theta_0}=\Delta_{\theta_x}=\Delta_{\theta_y}=0$ and this approach can reduce to the optimal strategy. This highlights the importance of incorporating appropriate prior knowledge for different contexts.

\stitle{Reducing computation complexity.}
Recall that the MAP strategy in \Eqref{eq:map} requires enumerating all $2^{N+M}$ combinations. Although the posterior probability is computable in \Eqref{eq:posterior_four_b_p1}, the enumeration cost still constrains the efficient identification of the optimal solution. Fortunately, given the role of the indicator $P_1$, only consistent combinations where $P_1=1$ need consideration. To be specific, for any $\hvx\in\{0,1\}^N$ and $\hvy\in\{0,1\}^M$ combination:
\begin{itemize}[leftmargin=*]
    \item $\hvx$ must conform to the consistency assumption (Assumption \ref{asm:identical_functionality}). Thus, any correct solution $i$ with $\hx_i=1$ must pass the same test cases, \ie, they should be within the same consensus set.
    \item $\hvy$ must match the test cases passable by any correct solution, meaning all correct test cases $j$ with $\hy_j=1$ should also reside in the corresponding consensus set of the correct solutions.
\end{itemize}

Therefore, we claim that valid combinations must ensure that \textit{all correct solutions and test cases should be in the same consensus set}. To reduce computations further, we consider any two solutions within the same consensus set. As these solutions pass identical test cases, they are completely symmetric and indistinguishable in $\mE$. Therefore, it is illogical to differentiate between them. Thus, we assume that \textit{solutions within the same consensus set should have identical predicted correctness}.

Based on these insights, we propose an enumeration method based on consensus sets. Similar to \method{CodeT}, we initially divide solutions and test cases into $K$ consensus sets $(S_i^x, S_i^y)_{i=1}^K$. Within each set $(S_i^x, S_i^y)$, we predict all solutions in $S_i^x$ as 1 and all test cases in $S_i^y$ as 1, while others are predicted as 0. This forms a consistent configuration $(\hvx, \hvy)$. We then calculate the posterior of $(\hvx, \hvy)$ with \Eqref{eq:posterior_four_b}, where $P_1=1$ is always satisfied. This significantly reduces the number of explored configurations from $2^{N+M}$ to $K$.

\subsection{Incorporating Prior Knowledge}\label{sec:prior}

We have derived a general explicit expression for the posterior probability in \Eqref{eq:posterior_four_b_p1}, which includes eight hyperparameters corresponding to the Beta distribution for four $\theta$. According to \cref{thm:bound_b4}, we should incorporate proper prior knowledge to effectively approximate the optimal strategy. In this section, we investigate how to achieve this in the context of code generation.

\stitle{Priors for $\theta_0$ and $\theta_1$.}
In practical scenarios, a test suite, not to mention a test case, is often incomplete. Therefore, a correct test case can fail to identify an incorrect solution, causing incorrect solutions to have a moderate probability of passing correct test cases (\ie, $\theta_1$). Conversely, to pass incorrect test cases that validate flawed functionalities, incorrect solutions must "accidentally" match this specific flaw to pass, making such occurrences ($\theta_0$) relatively rare. This suggests that in practice, $\theta_0$ may be very small, but $\theta_1$ may not have a clear pattern.

\begin{figure}[t]
  \centering
  \subfigure[Distributions for $\theta_0$ and $\theta_1$]{
  \label{fig:distribution_theta_01}
    \includegraphics[width=0.22\textwidth]{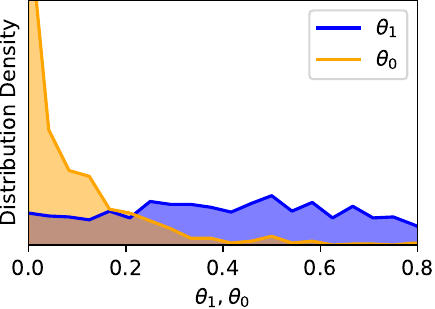}
  }
  \subfigure[Beta distributions]{
  \label{fig:distribution_beta}
    \includegraphics[width=0.22\textwidth]{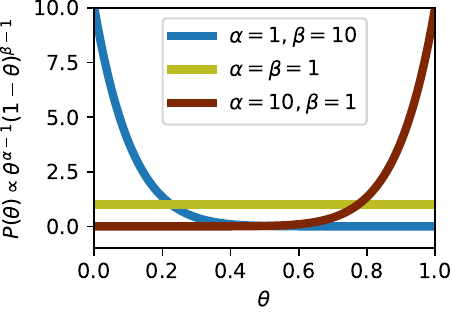}
  }
  \vspace{-4mm}
  \caption{(a) Real distributions for two parameters $\theta_0$ and $\theta_1$. (b) Three Beta distributions with different hyperparameters.}
  \label{fig:beta_visual}
\end{figure}

To validate this conjecture, we analyzed code and test case generation tasks with five different models on HumanEval (See \cref{sec:exp_setup} for details of models) and computed the actual values of $\theta_1$ and $\theta_0$ for each problem in HumanEval using ground-truth solutions. \cref{fig:distribution_theta_01} displays the true distributions of these parameters, showing that most $\theta_0$ values are concentrated near zero, while $\theta_1$ tends to follow a uniform distribution. 

Based on this finding, we propose adopting a prior distribution approaching zero for $\theta_0$ and a uniform prior distribution for $\theta_1$. Therefore, we choose a beta prior distribution parameterized by $(\alpha_0=1, \beta_0 \gg 1)$ for $\theta_0$, and choose $(\alpha_1=\beta_1=1)$ for $\theta_1$. As demonstrated in \cref{fig:distribution_beta}, such choice aligns with the findings in \cref{fig:distribution_theta_01}. In practice, $\beta_0$ serves as a tunable hyperparameter.

\stitle{Priors for $\theta_x$ and $\theta_y$.} As discussed previously, each consistent $(\hvx, \hvy)$ corresponds to a consensus set. Chen \etal~\cite{chen2023codet} identified a heuristic rule that the consensus set with the largest capacity (\ie, $n_x n_y$) is most likely correct. We will validate this rule theoretically in \cref{sec:theorem}. Accordingly, we want the prior distribution $p(\hvx, \hvy)$ to favor configurations containing more ones and reward larger consensus sets. This can be implemented by setting the hyperparameters for $\theta_x$ as $(\alpha_x\gg 1, \beta_x=1)$, and for $\theta_y$ as $(\alpha_y\gg 1, \beta_y=1)$, as illustrated in \cref{fig:distribution_beta}. Moreover, we find it sufficient to combine $\alpha_x$ and $\alpha_y$ into a single hyperparameter $\alpha_{xy}$, further reducing the parameter tuning space (see Section \ref{sec:exp_ablation_xy} for details).


\vspace{1.5mm}

\begin{mdframed}[linecolor=black,linewidth=1pt]
\vspace{-0.5mm}
\textbf{Answer to \textit{RQ3}: } 
A practical strategy to approximate uncomputable optimal strategy is to score $K$ consensus sets and select solutions within the highest-score set. The score is determined by multiplying 4 Beta functions, \ie, 
\begin{equation}
\begin{aligned}
    &\betafunc{n_1+1, |\mE_1|-n_1+1}\cdot \betafunc{n_0+1, |\mE_0|-n_0+\beta_0}\\
    \label{eq:posterior_four_b}
    \cdot &\betafunc{n_x+\alpha_{xy}, N-n_x+1}\cdot \betafunc{n_y+\alpha_{xy}, M-n_y+1},
\end{aligned}
\end{equation}
where $\beta_0$ and $\alpha_{xy}$ are tunable hyperparameters.
\vspace{-0.5mm}
\end{mdframed}

\subsection{Further Analysis of Algorithm \ours}\label{sec:analysis}

Given that the score in \Eqref{eq:posterior_four_b} is multiplied by four \underline{B}eta functions, we name this practical strategy \ours. In this section, we provide a detailed analysis of the proposed \ours~to deepen the understanding.

\stitle{Full algorithm.}
Algorithm \ref{alg:b4} outlines the workflow. Line 1 starts by collecting the set of test cases each code $i$ passes (denoted as $\ve_i$, \ie, $\{e_{i1},\cdots,e_{iM}\}$) and removes duplicates. In Line 3, we iterate over all unique test case sets. For each $\hvy$ processed, we identify solutions whose passed test cases precisely match $\hvy$ as $\hvx$ in Line 4. Note that $\hvx$ and $\hvy$ define a consensus set together. Lines 5-9 compute the score of this consensus set (\ie, the posterior) by \Eqref{eq:posterior_four_b}. Ultimately, Lines 10-11 identify the consensus set with the highest score as the prediction. For numerical stability, we often store the logarithm of the scores in practice, by summing the logarithms of the four Beta functions.

\setlength{\belowdisplayskip}{2pt}
\setlength{\belowdisplayshortskip}{2pt}
\setlength{\textfloatsep}{4mm}

\begin{algorithm}[t]
\caption{Algorithm for \ours}
\label{alg:b4}
\KwIn{Passing matrix $\mE=\{e_{ij}\}\in\{0,1\}^{N\times M}$, hyperparameters $\beta_0>1, \alpha_{xy}>1$}
\KwOut{$\hvx^*\in\{0,1\}^N$ and $\hvy^*\in\{0,1\}^M$ indicating the predicted correct solutions and test cases}
$S^y\gets\method{Deduplicate}(\{\ve_i\mid i\in[N]\})$\;
$Score^*\gets -\infty$\;
\For{$\hvy \in S^y$}{
    $\hvx \gets\left\{
        \indicator{\ve_i=\hvy}\mid i\in[N]
    \right\}$\;
    $\mE_{1}\gets\{e_{ij}\mid \hx_i=0, \hy_j=1, i\in[N], j\in[M]\}$\;
    $\mE_{0}\gets\{e_{ij}\mid \hx_i=0, \hy_j=0, i\in[N], j\in[M]\}$\;
    $n_1\gets\sum_{e\in \mE_{1}} e$,\quad\quad\ $n_0\gets\sum_{e\in \mE_{0}} e$\;
    $n_x\gets \sum_{i\in [N]}\hx_i$, \quad $n_y\gets \sum_{j\in [M]}\hy_j$\;
    $Score\gets \betafunc{n_1+1,|\mE_1|-n_1+1}\cdot \betafunc{n_0+1,|\mE_0|-n_0+\beta_0} \cdot \betafunc{n_x+\alpha_{xy},N-n_x+1} \cdot \betafunc{n_y+\alpha_{xy},M-n_y+1}$\;
    \If{$Score>Score^*$}{
        $(Score^*, \hvx^*, \hvy^*) \gets (Score, \hvx, \hvy)$\;
    }
}
\Return{$\hvx^*$, $\hvy^*$}\;
\end{algorithm}

\stitle{A running example.} We reuse \cref{fig:example} to illustrate how \ours~works, using the hyperparameters $\beta_0=\alpha_{xy}=10$.
Firstly, we deduplicate the rows in \Eqref{eq:passing_matrix_example} and obtain $S^y=\{[1, 1, 1, $ $ 0, 0], [0, 1, 1, 1, 1], [0, 0, 1,$ $1, 0]\}$, indicating there are three distinct sets of passed test cases corresponding to three consensus sets. We need to iterate all three sets and score for each one. For the first iteration, $\hvy=[1, 1, 1, 0, 0]$ and $\hvx=[1, 1, 0, 0]$. It indicates the first consensus set is $(\{x_1,x_2\}, \{y_1,y_2,$ $y_3\})$. Using \Eqref{eq:e12}, we obtain:
\vspace{-1.5mm}
\begin{align*}
    \mE_1=&\{e_{ij}\mid \hx_i=0, \hy_j=1\}=\{e_{31}, e_{32}, e_{33}, e_{41}, e_{42}, e_{43}\},\\
    \mE_0=&\{e_{ij}\mid \hx_i=0, \hy_j=0\}=\{e_{34}, e_{35}, e_{44}, e_{45}\},
\end{align*}
where $\mE_1$ (or $\mE_0$) represents the events that an incorrect solution passes a correct (or an incorrect) test case, under the prediction $\hvx$ and $\hvy$. We count these events: $n_1=\sum \mE_1=3$, $n_0=\sum \mE_0=3, n_x=\sum \hvx=2$, and $n_y=\sum\hvy=3$. Following this, the score is:
\begin{align*}
    &\betafunc{3+1, 6-3+1}\times\betafunc{3+1, 4-3+10}\\
    &\times \betafunc{2 + 10, 4-2+1}\times\betafunc{3+10, 5-3+1}=\mathbf{1.20\times 10^{-12}}.
\end{align*}

For the second iteration, we have $\hvy=[0, 1, 1, 1, 1]$ and $\hvx=[0, 0, 1, 0]$, resulting the score $\mathbf{1.15\times 10^{-13}}$. For the third iteration, we have $\hvy=[0, 0, 1, 1, 0]$ and $\hvx=[0, 0, 0, 1]$, resulting the score $\mathbf{1.24\times 10^{-15}}$. One can find that the first consensus set has the largest score $1.20\times 10^{-12}$, leading to the selection of $\{x_1, x_2\}$ as the optimal solution.

\begin{figure}[t]
  \centering
  \subfigure[$\log\betafunc{n_0+1,|\mE_0|-n_0+\beta_0}$]{
    \label{fig:beta_func_beta0}
    \includegraphics[width=0.22\textwidth]{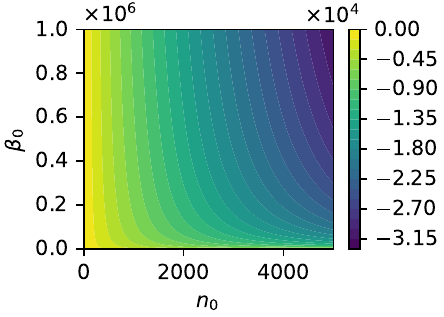}
  }
  \subfigure[$\log\betafunc{n_x+\alpha_{xy}, N-n_x+1}$]{
    \label{fig:beta_func_alpha}
    \includegraphics[width=0.22\textwidth]{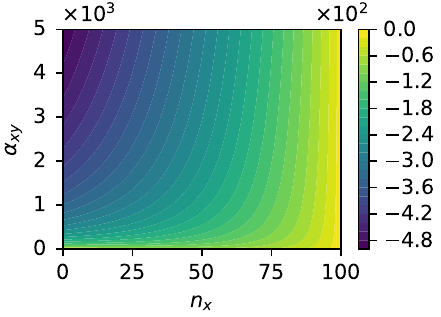}
  }
  \vspace{-4mm}
  \caption{Visualization of two Beta functions used in our scoring strategy. We set $|\mE_0|=5000$ and $N=100$.\\}
  \label{fig:beta_func}
\end{figure}

\stitle{Understanding Beta functions.} To further explore the role of two hyperparameters used in the \ours~and our scoring strategy, we visualize two Beta functions related to two hyperparameters $\beta_0$ and $\alpha_{xy}$ in \cref{fig:beta_func}. \cref{fig:beta_func_beta0} reveals that the function value is insensitive to $n_0$ when $\beta_0$ is very small. As $\beta_0$ increases, the Beta function has little change for small $n_0$ but has a particularly small value for large $n_0$. This suggests that a larger $\beta_0$ leads the algorithm to reward predictions with smaller $n_0$. Recall that $n_0$ represents the number of incorrect solutions passing incorrect test cases, which is generally small in the real world (as discussed in \cref{sec:prior}). This indicates that our \ours, which uses a $\beta_0 \gg 1$, aligns with practical conditions well. Similarly, \cref{fig:beta_func_alpha} shows a large $\alpha_{xy}$ leads the algorithm to predict more correct solutions or tests (\ie, larger $n_x$ or $n_y$), which rewards a larger consensus set as we expected in \cref{sec:prior}.

\;\\

\section{Theoretical analysis}\label{sec:theorem}

In this section, we address \textit{RQ4} by a theoretical accuracy analysis of the two representative heuristics, \method{MaxPass} and \method{CodeT}, to investigate under what conditions they can and cannot work. \method{MaxPass} is a widely-used heuristic \cite{lahiri2023interactive,li2022competition,le2022coderl,roziere2022leveraging} and \method{CodeT} is the state-of-the-art heuristic for code generation. Furthermore, these theoretical analyses further explain why the priors for $P(\hvx, \hvy)$ introduced in Section \ref{sec:prior} are chosen. We assume that Assumptions \ref{asm:identical_functionality}-\ref{asm:dependence_e} are satisfied, and the data follows the generation process in \cref{fig:intro-method}. All proofs can be found in the online Appendix \cite{zenodo}. 

We begin with a theorem which assesses \method{MaxPass}'s accuracy when there is a large number of test cases:

\begin{lemma}\label{thm:maxpass}
    Suppose there exist $n_y$ correct test cases and $\nny$ incorrect test cases ($n_y+\nny=M$). When both $n_y$ and $\nny$ are large enough, the probability of any incorrect code passing $Y$ ($Y\geq n_y$) test cases is:
    \begin{align*}
        P(Y\geq n_y)\sim \Phi\left(
            \frac{
                \nny\theta_0-n_y(1-\theta_1)
            }{
                \sqrt{n_y \theta_1(1-\theta_1)+\nny\theta_0(1-\theta_0)}
            }
        \right),
    \end{align*}
    where $\Phi$ is the cumulative distribution function (CDF) of the standard normal distribution. $\theta_0$ and $\theta_1$ are defined in \Eqref{eq:eij}.
\end{lemma}
\begin{theorem}[Impact of correct test cases for \method{MaxPass}]
    \label{thm:maxpass_ny}
    If $\theta_1<1$, the accuracy of \method{MaxPass} (\ie, the probability of all incorrect solutions passing less than $n_y$ test cases) can exponentially converge to $1$ as $n_y\rightarrow\infty$.
\end{theorem}
\begin{theorem}[Impact of incorrect solutions for \method{MaxPass}]
    \label{thm:maxpass_nx}
    If there are $\nnx$ incorrect solutions, the accuracy of \method{MaxPass} can exponentially converge to $0$ as $\nnx\rightarrow\infty$.
\end{theorem}

\begin{figure*}[t]
  \centering
  \subfigure[Varying $N$]{
    \label{fig:exp_sim_n}
    \includegraphics[width=0.24\textwidth]{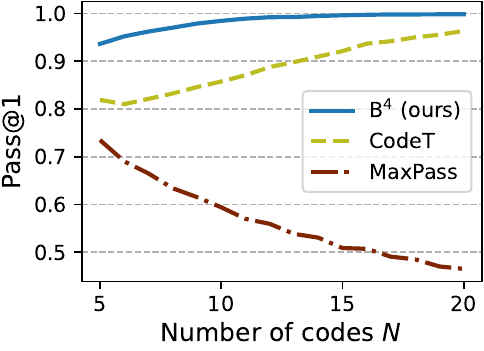}
  }%
  \subfigure[Varying $M$]{
    \label{fig:exp_sim_m}
    \includegraphics[width=0.24\textwidth]{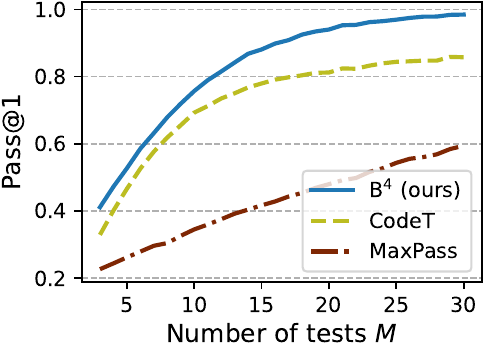}
  }%
  \subfigure[Varying $\theta_x$]{
    \label{fig:exp_sim_x}
    \includegraphics[width=0.24\textwidth]{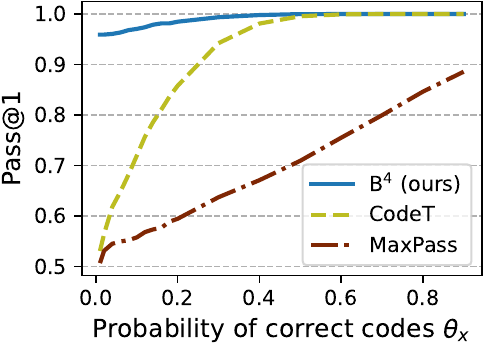}
  }%
  \subfigure[Varying $\theta_y$]{
    \label{fig:exp_sim_y}
    \includegraphics[width=0.24\textwidth]{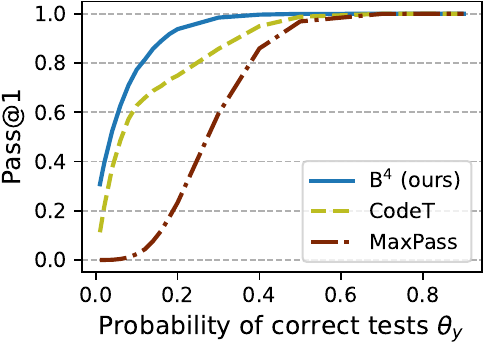}
  }
  \vspace{-4mm}
  \caption{Pass@1 results of the three methods under different conditions in the simulated experiments. By default, we set $N=10$, $M=30$, $\theta_x=0.2$, and $\theta_y=0.3$ except for the varied one.}
  \label{fig:exp_simulate}
\vspace{-3mm}
\end{figure*}

\cref{thm:maxpass_ny} demonstrates the working condition for \method{MaxPass}: it requires a large amount of correct test cases $n_y$ to make the accuracy converge to 1. However, \cref{thm:maxpass_nx} also underscores a limitation of \method{MaxPass}: it lacks \textit{scalability} to the number of code solutions $N$. As $N$ increases, $\nnx$ increases and the accuracy of \method{MaxPass} will exponentially converge to zero.


Following this, we analyze the error of \method{CodeT}.
Considering the problem's complexity, we \textit{fix} the $M$ test cases and explore how the error evolves as the number of generated code solutions $N$ grows, as shown in the following theorem.

\begin{lemma}\label{thm:codet}
    Suppose the correctness of code solutions and test cases are $\vx$ and $\vy$. Let $n_x=\sum \vx$ and $n_y=\sum \vy$ denote the number of correct code solutions and test cases, respectively. For any \textbf{incorrect} consensus set that corresponds to a prediction $\hvx$ and $\hvy$, similarly let $n_{\hat x}=\sum \hvx$ and $n_{\hat y}=\sum \hvy$. For arbitrary $\vy$ and $\hvy$, if $N$ is sufficiently large, the probability of this consensus set being scored higher than the correct one by \method{CodeT} (\ie, $n_{\hat x} n_{\hat y} > n_x n_y$) follows:
    {\small
    \begin{align*}
        P(n_{\hat x}  n_{\hat y} > n_x n_y)
        \sim \Phi\left(
            \frac{
                \sqrt{N} (\theta' n_{\hat y} - \theta_xn_y)
            }{
                \sqrt{n_{\hat y}^2 \theta'(1-\theta')+n_y^2 \theta_x(1-\theta_x)-2n_{\hat y}n_y \theta' \theta_x}
            }
        \right),
    \end{align*}
    }
    where $\theta'$ is a constant, defined as:
    \begin{align*}
        \theta' = (1-\theta_x) \theta_1^{\hvy^\top \vy} (1-\theta_1)^{(1-\hvy)^\top\vy}{\theta_0}^{\hvy^\top (1-\vy)}(1-\theta_0)^{(1-\hvy)^\top (1-\vy)}.
    \end{align*}
\end{lemma}
\begin{theorem}[Impact of $\theta_x$ and $N$ for \method{CodeT}]
    \label{thm:codet_error_converge}
    If $\theta_x$ is large enough such that $\theta' n_{\hat y} < \theta_x n_y$, then the error probability $P(n_{\hat x}  n_{\hat y} > n_x n_y)$ can exponentially converge to 0 as $N\rightarrow\infty$. Otherwise, if $\theta_x$ is low enough such that $\theta' n_{\hat y} > \theta_x n_y$, the error probability converges to 1 as $N\rightarrow\infty$.
\end{theorem}

\cref{thm:codet_error_converge} elucidates the working condition for \method{CodeT}: it requires a sufficient high correct probability of code solutions (high $\theta_x$). If the generated solutions contain excessive incorrect solutions, \method{CodeT} may not work well. An important insight is that under the condition of high $\theta_x$, \method{CodeT} offers better scalability compared to \method{MaxPass}: as the number of solutions $N$ increases, \method{CodeT}'s selection accuracy can exponentially converge towards 1 (\cref{thm:codet_error_converge}), whereas MaxPass's accuracy will converge towards 0 (\cref{thm:maxpass_nx}). 

\vspace{3mm}

\begin{mdframed}[linecolor=black,linewidth=1pt]
\vspace{-0.5mm}
\textbf{Answer to \textit{RQ4}: } Existing heuristics work under specific conditions. \method{MaxPass} requires sufficient correct test cases, while \method{CodeT} requires a high correct probability of solutions. When both of their requirements are satisfied, \method{CodeT} has better scalability with the number of solutions $N$ than \method{MaxPass}. 
\vspace{-0.5mm}
\end{mdframed}

\vspace{3mm}

Considering the analyzing complexity, whether a similar error probability analysis can be directly provided for \ours~is an open question.\footnote{To show the complexity, note that computing the distribution for \ours's score is necessary for estimating error probability. The score can be represented as the product of $n_x$, $n_1$, and $n_0$ after nonlinear transformations (here we assume $n_y$ is given, as \cref{thm:codet}). However, despite oversimplification, \ie, treating three variables as normal, linearizing the transformations, and assuming their independence, the computation is still a challenge in the literature \cite{stojanac2017products}.} Fortunately, these theoretical analyses still indirectly support the effectiveness of \ours. For example, \cref{thm:codet_error_converge} validates the effectiveness of the priors for $\theta_x$ and $\theta_y$ of our \ours. Recall that our introduced priors for $P(\hvx, \hvy)$ are similar to \method{CodeT}'s assumptions (\cref{sec:prior}), which offers similar scalability benefits under the condition that $\theta_x$ is relatively large. However, it is crucial to note that these priors are just part of our methods. Besides the priors for $\theta_x$ and $\theta_y$, we also incorporate priors for $\theta_0$ and $\theta_1$, which effectively compensates for the limitations of \method{CodeT}’s priors, particularly in scenarios where $\theta_x$ is low. As our subsequent experiments confirm, \ours~significantly outperforms \method{CodeT} in such challenging scenarios.


\section{Experiment}\label{sec:exp}

In this section, we conduct experiments to further answer \textit{RQ4} and \textit{RQ5}. We start with exploring the conditions under which existing heuristics can work efficiently through simulation experiments in different controlled environments, to validate the theoretical insights discussed in \cref{sec:theorem}. Subsequently, we compare the performance of \ours~with existing heuristics on real-world datasets.

\subsection{Simulated Experiments}\label{sec:exp_simulated}

In our simulated experiments, we sampled $N=10$ solutions and $M=30$ test cases, and set four parameters $\theta_x=0.2$, $\theta_y=0.3$, $\theta_1=0.4$ and $\theta_0=0.1$ by default. These default values are based on our measurement of the real data generated by CodeGen \cite{nijkamp2023codegen} on HumanEval \cite{chen2021evaluating}. Based on these parameters, we randomly sampled a data point $(\vx, \vy, \mE)$ following the process shown in \cref{fig:intro-method}. Subsequently, we used \method{MaxPass}, \method{CodeT}, and \ours~to select the solutions $\hvx$ using $\mE$, and computed the proportion of correct solutions within $\hvx$ (\ie, Pass@1) using the ground-truth $\vx$. We repeated this process 20,000 times and averaged the results to ensure stability for each experiment. Following \cref{sec:prior}, the hyperparameters $\beta_0$ and $\alpha_{xy}$ should be larger than 1, and we preliminarily chose $\beta_0=\alpha_{xy}=10$.

\cref{fig:exp_sim_n,fig:exp_sim_m} display the results as the scale of data $N$ and $M$ change. One can observe in \cref{fig:exp_sim_n} that \method{CodeT}'s performance gradually improves with an increase in the number of code solutions $N$, whereas \method{MaxPass} shows a decline as $N$ increases. This confirms our theoretical results in Section \ref{sec:theorem}: \method{CodeT} has better scalability with $N$ than \method{MaxPass}. \cref{fig:exp_sim_m} shows that unlike with $N$, \method{MaxPass} tends to improve as $M$ increases. Regardless of the values of $N$ and $M$, \ours~consistently outperforms the two baselines, proving that existing heuristic algorithms are not optimal. Specifically, \ours~tends to provide greater performance enhancements relative to \method{CodeT} when $N$ is small. This could be because \method{CodeT} does not perform as well when $N$ is low, which is also validated in \cref{thm:codet_error_converge}.

\cref{fig:exp_sim_x,fig:exp_sim_y} display the results as the probability of correct solutions $\theta_x$ and test cases $\theta_y$ change. All three methods gradually improve as the accuracy increases. Specifically, both \ours~and \method{CodeT}'s accuracies can converge to 1 as $\theta_x$ increases, while all three methods converge to 1 as $\theta_y$ increases. This indicates that \method{MaxPass} is less sensitive to $\theta_x$ but more responsive to $\theta_y$, confirming the findings of \cref{thm:maxpass} that the number of correct test cases matters for \method{MaxPass}.  \ours~consistently outperforms all the two heuristics under all conditions. Notably, when $\theta_x$ is low, it significantly outperforms \method{CodeT} with a large improvement. This suggests that \method{CodeT} struggles under the condition of few correct solutions and affirms the findings of \cref{thm:codet_error_converge}. 

\subsection{Real-world Experiments}\label{sec:exp_realworld}

\subsubsection{Experiment setup}\label{sec:exp_setup}

We conducted experiments on three public code generation benchmarks, HumanEval \cite{chen2021evaluating}, MBPP \cite{austin2021program} (sanitized version), and APPS \cite{hendrycks2021measuring} with three difficulty levels. \added{These benchmarks have been widely used by LLM-based code generation studies \cite{chen2021evaluating,nijkamp2023codegen,li2023starcoder,rozière2024code,guo2024deepseek}. Specifically, each benchmark contains some coding tasks, and each task consists of a natural language requirement, a function signature, and a golden test suite for evaluating the correctness of generated solutions. Notably, these golden test suites and the generated test cases are not the same; the generated test cases are used by each selection strategy to select the generated code, while the golden test suites are solely used to evaluate the performance of selection strategies.}

We used the same zero-shot prompt format as \method{CodeT}~\cite{chen2023codet} for both code and test case generation. Following \method{CodeT}, the numbers of generated solutions and test cases are 100 for HumanEval and MBPP and 50 for APPS. Both solutions and tests are generated by the same model.

For models, our experiments are based on Codex \cite{chen2021evaluating} (code-davinci-002 version), CodeGen \cite{nijkamp2023codegen} (16B Python mono-lingual version), and three recent open-source models, StarCoder \cite{li2023starcoder}, CodeLlama \cite{rozière2024code} (7B Python version) and Deepseek-Coder \cite{guo2024deepseek} (6.7B Instruct version). The generation hyperparameters such as temperature, top $p$, and max generation length are the same as \cite{chen2023codet}. Additionally, as APPS has significantly more problems (5,000) compared to HumanEval (164) and MBPP (427), testing all models on it is prohibitively expensive. Given that Codex outperforms the other models on HumanEval and MBPP in most of our experiments (using \method{CodeT} strategy), we followed Chen \etal~\cite{chen2023codet} by only evaluating Codex's outputs on the APPS dataset.


For baselines, in addition to \method{MaxPass} \cite{lahiri2023interactive,le2022coderl} and \method{CodeT} \cite{chen2023codet}, we also used \method{MBR-exec} \cite{shi2022natural,li2022competition}, which is similar to \method{CodeT} but scores each consensus set with the number of solutions, and a naive \method{Random}, which picks a code from the generated solutions randomly. We reported the average Pass@1 of the selected solutions. Our method is presented in the format of \ours($\log_{10} \beta_0$, $\log_{10} \alpha_{xy}$). For example, \ours(4,3) represents $\beta_0=10^4$ and $\alpha_{xy}=10^3$. For a fair comparison, all the methods operate on the same passing matrices $\mE$. We reported three variants of methods: \ours(4,3), \ours(5,3), and \ours(6,3), and compared each of them with \method{CodeT} using Wilcoxon signed-rank significance test \cite{wilcoxon1992individual}.

\begin{table*}[t]
  \centering
  \caption{Pass@1 (\%) of the code solutions selected by different strategies across various datasets and models with two settings (RD=\method{Random}, MP=\method{MaxPass}, MBR=MBR-\method{exec}, CT=\method{CodeT}). We also reported the average relative improvement of the three \ours~variants over the strongest heuristic \method{CodeT} and the p-values derived from the Wilcoxon signed-rank test. }
  \vspace{-2mm}
  \resizebox{\linewidth}{!}{
\begin{tabular}{llccccccccccccccc}
\toprule
{\multirow{3}[6]{*}{\textbf{Dataset}}} & {\multirow{3}[6]{*}{\textbf{Model\;\;\;\;\;\;\;\;\;\;\;\;\;\;\;\;\;\;\;\;\;\;\;\;}}} & \multicolumn{7}{c}{\boldmath{}\textbf{Discriminative Problems (0 < $\theta_x$ < 1)}\unboldmath{}} &  \;\;\;\;\;\;    & \multicolumn{7}{c}{\boldmath{}\textbf{Hard Problems (0 < $\theta_x$ < 0.5)}\unboldmath{}} \\
\cmidrule{3-9}\cmidrule{11-17}     &      & {\multirow{2}[4]{*}{\textbf{RD}}} & {\multirow{2}[4]{*}{\textbf{MP}}} & {\multirow{2}[4]{*}{\textbf{MBR}}} & {\multirow{2}[4]{*}{\textbf{CT}}} & \multicolumn{3}{c}{\textbf{ours}} &      & {\multirow{2}[4]{*}{\textbf{RD}}} & {\multirow{2}[4]{*}{\textbf{MP}}} & {\multirow{2}[4]{*}{\textbf{MBR}}} & {\multirow{2}[4]{*}{\textbf{CT}}} & \multicolumn{3}{c}{\textbf{ours}} \\
\arrayrulecolor{gray!80}
\cmidrule{7-9}\cmidrule{15-17}   
\arrayrulecolor{black}  
&      &      &      &      &      & \textbf{\ours(4,3)} & \textbf{\ours(5,3)} & \textbf{\ours(6,3)} &      &      &      &      &      & \textbf{\ours(4,3)} & \textbf{\ours(5,3)} & \textbf{\ours(6,3)} \\
\midrule
{\multirow{5}[2]{*}{HumanEval}} & {CodeGen} & {32.5} & {28.6} & {44.8} & {51.5} & 56.8  & \textbf{58.0} & 56.9  &      & {13.0} & {11.1} & {21.0} & {31.4} & 38.2  & 40.0  & \textbf{40.8} \\
     & {Codex} & {39.2} & {57.8} & {55.0} & {71.7} & 70.6  & \textbf{73.1} & \textbf{73.1} &      & {19.2} & {43.2} & {27.6} & {54.6} & 52.9  & \textbf{56.9} & \textbf{56.9} \\
     & {StarCoder} & {29.8} & {32.2} & {47.9} & {55.0} & 59.0  & \textbf{59.3} & 57.8  &      & {15.0} & {16.3} & {29.3} & {38.9} & 44.4  & \textbf{44.8} & 42.8  \\
     & {CodeLlama} & {34.1} & {40.6} & {52.6} & {61.7} & 63.5  & \textbf{64.8} & 64.0  &      & {15.8} & {24.5} & {30.8} & {44.1} & 46.7  & \textbf{48.6} & 47.4  \\
     & {Deepseek-Coder} & {65.3} & {58.2} & {80.4} & {79.2} & \textbf{80.5} & 78.5  & 78.5  &      & {24.7} & {33.7} & {35.0} & {30.6} & \textbf{35.5} & 31.3  & 31.3  \\
\arrayrulecolor{gray!80}
\midrule
\arrayrulecolor{black}
{\multirow{5}[1]{*}{MBPP}} & {CodeGen} & {42.4} & {48.1} & {56.4} & {64.9} & \textbf{66.7} & 64.9  & 64.7  &      & {21.8} & {30.8} & {28.4} & {43.5} & \textbf{45.6} & 42.5  & 42.3  \\
     & {Codex} & {55.1} & {70.5} & {71.9} & {80.0} & 80.8  & 81.3  & \textbf{81.9} &      & {23.9} & {46.4} & {32.5} & {53.9} & 55.1  & 56.6  & \textbf{58.0} \\
     & {StarCoder} & {46.1} & {55.6} & {65.6} & {69.6} & \textbf{70.6} & \textbf{70.6} & \textbf{70.6} &      & {21.5} & {39.5} & {37.8} & {45.6} & 47.5  & \textbf{47.9} & \textbf{47.9} \\
     & {CodeLlama} & {47.2} & {60.0} & {65.4} & {72.4} & 72.6  & 73.4  & \textbf{73.8} &      & {19.8} & {39.0} & {30.7} & {44.7} & 45.0  & 46.7  & \textbf{47.5} \\
     & {Deepseek-Coder} & {56.5} & {71.4} & {66.9} & {75.2} & \textbf{75.9} & \textbf{75.9} & \textbf{75.9} &      & {22.3} & {45.6} & {25.7} & {45.9} & 46.7  & \textbf{47.6} & \textbf{47.6} \\
\arrayrulecolor{gray!80}
\midrule
\arrayrulecolor{black}
{APPS introductory} & {\multirow{3}[1]{*}{Codex}} & {36.2} & {46.4} & {41.6} & {59.5} & 63.7  & 63.7  & \textbf{64.4} &      & {17.6} & {29.5} & {15.9} & {41.6} & 46.6  & 47.6  & \textbf{48.3} \\
{APPS interview} &      & {15.6} & {26.0} & {14.7} & {36.0} & 40.4  & 40.8  & \textbf{41.1} &      & {11.2} & {22.4} & {8.0} & {30.6} & 35.1  & 35.5  & \textbf{35.9} \\
{APPS competition} &      & {7.9} & {16.8} & {3.1} & {17.3} & 23.1  & \textbf{25.2} & \textbf{25.2} &      & {7.0} & {16.2} & {2.5} & {16.0} & 21.9  & \textbf{24.0} & \textbf{24.0} \\
\midrule
     \multicolumn{6}{r}{\textit{Avg. relative improvement over the strongest heuristic \method{CodeT}}} & +6.1\% & \textbf{+7.5\%} & +7.2\% &      &      &      &      &      & +10.1\% & \textbf{+12.0\%} & \textbf{+12.0\%} \\
& \multicolumn{5}{r}{\textit{p-value}} & 0.001 & 0.0003 & 0.0006 &      &      &      &      &      & 0.0009 & 0.0004 & 0.0004 \\
\bottomrule
\end{tabular}%
}
  \label{tab:main}
\vspace{-3mm}
\end{table*}%

To comprehensively evaluate different selection methods, we filtered the problems based on the proportion of correct solutions among all generated solutions (\ie, $\theta_x$). We first filtered out problems with $\theta_x=1$ and $\theta_x=0$, as the solutions for these problems are either entirely correct or incorrect, which can not differentiate selection strategies. We name this setting \textbf{discriminative problems}. To provide a more challenging environment for selecting correct solutions, we propose a new setting on a subset of discriminative problems where $0<\theta_x<0.5$, named \textbf{hard problems}.


\subsubsection{Main results}

\cref{tab:main} presents the main results, showing that all three \ours~variants consistently and significantly outperform existing heuristics. Specifically, each single variant of B4 outperforms all baselines in most cases. On average, each variant surpasses the strongest heuristic baseline, \method{CodeT}, by 6-12\% with statistically significant differences (proven by significance tests). This highlights a substantial gap between existing heuristics and the optimal strategy and suggests our method effectively approximates the optimal.

Additionally, \ours~shows a greater performance improvement over \method{CodeT} in more challenging scenarios (\ie, smaller $\theta_x$). It achieves a 6.1\%-7.5\% relative improvement in discriminative problems and a 10.1\%-12.0\% improvement in hard problems. In the most challenging scenario (APPS competition on hard problems), \ours~can even deliver up to a 50\% enhancement over \method{CodeT} and 246\% over random selection. These findings align with the conclusions of \cref{thm:codet} and the simulated experiments depicted in \cref{fig:exp_sim_x}, confirming that existing heuristics struggle with more difficult tasks. We also observed that the gains from \ours~on the MBPP dataset are less significant than on HumanEval and APPS, likely because the MBPP problems are inherently simpler, as indicated by \method{Random}.

For hyperparameters, the optimal hyperparameter $\beta_0$ for \ours~varies across different scenarios, suggesting that the prior distribution of $\theta_0$ may differ depending on the context. This makes sense as different models might generate incorrect solutions and test cases with different patterns. For example, when models more easily misinterpret the problem, leading solutions and test cases to follow the same incorrect patterns, the probability of incorrect solutions passing incorrect test cases $\theta_0$ can increase, thus necessitating a larger $\beta_0$ to reflect this change. We will further discuss the impact of hyperparameters in the next section.

\vspace{3mm}
\begin{mdframed}[linecolor=black,linewidth=1pt]
\vspace{-0.5mm}
\textbf{Answer to \textit{RQ5}: } The proposed \ours~significantly outperforms existing heuristics, achieving a 6.1\%-7.5\% relative improvement in discriminative problems and a 10.1\%-12.0\% improvement in hard problems over the strongest \method{CodeT}.
\vspace{-0.5mm}
\end{mdframed}

\begin{figure*}[t]
  \centering
  \hspace{-2mm}%
  \subfigure[HumanEval]{
    \label{fig:exp-param-humaneval}
    \includegraphics[width=0.33\textwidth]{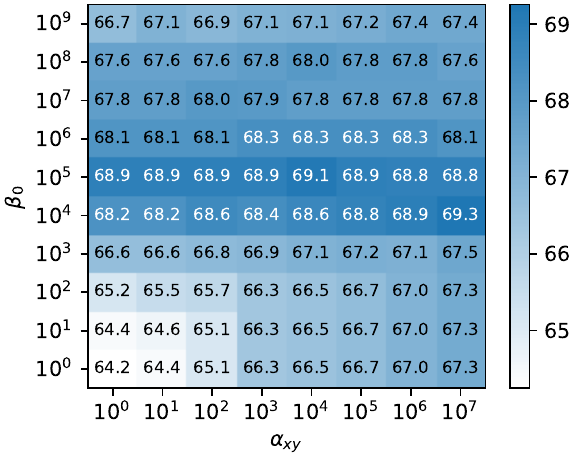}
  }%
  \hspace{-2mm}%
  \subfigure[MBPP]{
    \label{fig:exp-param-mbpp}
    \includegraphics[width=0.33\textwidth]{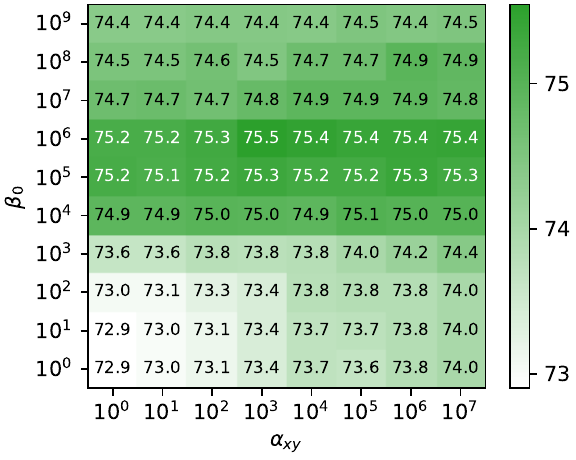}
  }%
  \hspace{-2mm}%
  \subfigure[\added{APPS}]{
    \label{fig:exp-param-apps}
    \includegraphics[width=0.33\textwidth]{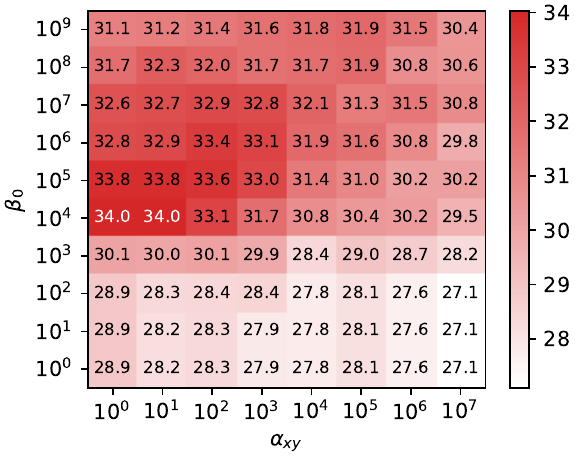}
  }
  \caption{Pass@1 (\%) results of varying $\alpha_{xy}$ and $\beta_0$ on HumanEval's, MBPP's, \added{and APPS' discriminative problems.}}
  \label{fig:exp_param}
\end{figure*}

\subsubsection{Ablation studies on two hyperparameters}\label{sec:exp_ablation_hyper}

\cref{fig:exp-param-humaneval,fig:exp-param-mbpp} show the average performance on two datasets as influenced by two hyperparameters $\beta_0$ and $\alpha_{xy}$. Recall that $\beta_0$ controls the likelihood $P(\mE\mid \hvx, \hvy)$ and $\alpha_{xy}$ controls the prior $P(\hvx, \hvy)$. For $\beta_0$, performance on both datasets initially increases and decreases as $\beta_0$ increases, with the optimal value around $10^4$-$10^6$. This pattern suggests that an appropriate $\beta_0$ can better align with the prior distribution of $\theta_0$, resulting in more accurate likelihood estimates. 

\added{For $\alpha_{xy}$, we found that performance improves with an increase in $\alpha_{xy}$ on HumanEval and MBPP, whereas the opposite is true for APPS. Recall that a larger $\alpha_{xy}$ makes the strategy closer to \method{CodeT}. One possible reason is that the tasks in HumanEval and MBPP are relatively simpler, so \method{CodeT} performs better on these two datasets, as shown in \cref{thm:codet_error_converge}.}


\begin{figure}[t]
  \centering
  \label{fig:exp-param-xy}
  \includegraphics[width=0.33\textwidth]{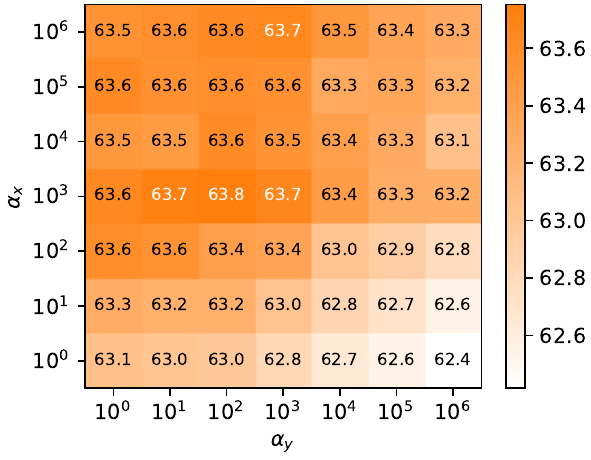}
  \caption{Pass@1 (\%) results of splitting $\alpha_{xy}$ into two hyperparameters $\alpha_x$ and $\alpha_y$ on the discriminative problems of HumanEval and MBPP when $\beta_0=10^6$.}
\end{figure}

\subsubsection{Ablation studies on splitting $\alpha_{xy}$ into two individual hyperparameters $\alpha_x$ and $\alpha_y$} \label{sec:exp_ablation_xy}

As discussed in \cref{sec:prior}, we combined $\alpha_x$ and $\alpha_y$ into a single $\alpha_{xy}$ in \Eqref{eq:theta_prior}. This section examines the effects of tuning $\alpha_x$ and $\alpha_y$ independently. \cref{fig:exp-param-xy} shows the trend of average performance across all datasets as $\alpha_x$ and $\alpha_y$ vary, with $\beta_0$ set at $10^6$. We observe that performance declines significantly when $\alpha_y-\alpha_x$ has a large value (\ie, in the bottom right area of \cref{fig:exp-param-xy}). As $\alpha_y-\alpha_x$ gradually decreases (moving from the bottom right towards the top left), performance can be gradually improved. The method achieves optimal performance when $\alpha_x$ and $\alpha_y$ are closed ($\alpha_x=10^3$ and $\alpha_y=10^2$). Considering that the model's performance is not sensitive to $\alpha_{xy}$ when $\beta_0$ is within an appropriate range, we argue that merging $\alpha_x$ and $\alpha_y$ into one hyperparameter simplifies tuning without substantially affecting performance. Therefore, we adopted $\alpha_x=\alpha_y=10^3$ in our previous main experiment.

\subsubsection{Computational Cost}

\added{\cref{tab:computation} shows the running time of the \ours~algorithm and \method{CodeT}, where \ours~is slightly slower than \method{CodeT} due to the relatively higher overhead of beta functions in \ours~compared to simple counting in \method{CodeT}. Notably, the computational complexity of both is the same, as both first partition the consensus sets and then score them. We can observe that even for large $M$ and $N$ (\eg, $M=N=400$), the running time is less than one second, which is much less than the time to generate 400 solutions and tests with LLMs. Therefore, we believe that the efficiency of \ours~will not become a bottleneck for practical systems. }


\section{Discussion}\label{sec:discussion}
In this section, we discuss the limitations and threats to the validity of this study.

\subsection{Limitations}


\textbf{\cref{asm:independence_xy} and \ref{asm:dependence_e}.} These assumptions are related to independence. \cref{asm:independence_xy} considers the correctness of code solutions and test cases are independent, which can be violated if there is a causal relationship in their generation, such as using a generated test case as input to an LLM for further generation. \cref{asm:dependence_e} states that passing probability is solely determined by the correctness of the associated code and test case. However, the independence of passing states may be broken by other unobserved factors hidden in the code. For example, if two incorrect solutions exhibit similar structures and similar error types, their passing states might be positively correlated. Considering the significant complexity introduced by the lack of independence, further exploration of the dependence case is deferred to future research.

\begin{table}[t]
    \centering
    \caption{\added{Computation cost with the increases of the number of code solutions $N$ and test cases $M$.}}
    \label{tab:computation}
    \begin{tabular}{ccccc}
        \toprule
        $N$ and $M$ & 100 & 200 & 300 & 400 \\
        \midrule
        \method{CodeT} & 10 ms & 65 ms & 202 ms & 455 ms \\
        \ours & 15 ms & 79 ms & 243 ms & 588 ms\\
        \bottomrule
    \end{tabular}
\end{table}


\textbf{Prior for $\theta_0$.} This prior assumes that $\theta_0$ (\ie, the probability of incorrect solutions passing incorrect test cases) is typically low. However, when LLMs misinterpret a problem, incorrect test cases may coincidently specify the functionality of incorrect solutions and potentially increase $\theta_0$. Considering that this prior can bring considerable benefits (as shown in \cref{sec:exp_ablation_hyper}), we argue that its advantages significantly outweigh the limitations.

\textbf{Priors for $\theta_x$ and $\theta_y$.} These priors, similar to the heuristic rule of \method{CodeT}, suggest that larger consensus sets are more likely to be correct. We have validated its theoretical effectiveness under the conditions of large $N$ and high $\theta_x$, as detailed in \cref{sec:theorem}. Even though its efficacy may diminish when these conditions are not met, the prior for $\theta_0$ effectively compensates for this situation as demonstrated in \cref{sec:exp_ablation_hyper}.

\textbf{Hyperparameters.} Our method includes two hyperparameters, $\alpha_{xy}$ and $\beta_0$, which may pose challenges in tuning across different usage scenarios. Fortunately, we have found that using consistent hyperparameters across all benchmarks can still yield significant improvements in our experimental scenarios. The tuning of hyperparameters for specific applications, potentially using a validation set to optimize them, remains an area for future research.

\textbf{Theoretical results.} To derive a closed form of the probabilities, we used \textit{the Law of Large Numbers} to examine the scenarios where $N$ and $M$ are sufficiently large. Besides, in \cref{thm:codet}, we focus on a single incorrect consensus set and neglect the complex interactions of multiple incorrect sets for computational convenience. Despite these simplifications, the key insights from these theorems are empirically validated in \cref{sec:exp}, thus we believe these theoretical analyses remain valuable. Finally, whether an error probability of \ours~can be explicitly provided, similar to those of existing heuristics provided in \cref{sec:theorem}, is an interesting open question.

\subsection{Threats to Validity}
The used benchmarks, \ie, HumanEval, MBPP, and APPS, consist of small-scale function-level tasks and may not capture the nuances of more complex scenarios in practice. Additionally, some ground-truth test suites used to evaluate the solution's correctness in the benchmarks are just an approximation to the specification and can be incomplete. This leads to a few correct solutions (\ie, the solutions passing the ground truth test suite) not exhibiting identical functionality and violating \cref{asm:identical_functionality}. Considering that such cases are relatively rare and most related work is centered on these benchmarks \cite{chen2023codet,chen2021evaluating,rozière2024code,li2023starcoder,guo2024deepseek}, we believe this threat will not significantly influence our conclusions. 

Our experiments focus on Python code generation tasks, which may not reflect the effectiveness of our method on other programming languages and other software engineering (SE) generation tasks. However, Python is one of the most popular programming languages and code generation is a challenging and important SE generation task. In addition, our method is language-agnostic and our theoretical framework can be easily adapted to other SE generation tasks, such as Automated Program Repair (APR) and code translation. Therefore, we believe this threat is limited. 

\section{Related work}

\stitle{Reranking and selection for plausible solutions.}
Using external validators (\eg, test cases) to assess, rerank, or select the generated solutions is widely used in various software engineering tasks. In code generation, Lahiri \etal~\cite{lahiri2023interactive} incorporated user feedback to choose test cases for code selection. 
In APR, Yang \etal~\cite{yang2017better} used test cases generated by fuzz testing to validate automatically generated patches. In code translation, Roziere \etal~\cite{roziere2022leveraging} leveraged EvoSuite \cite{fraser2011evosuite} to automatically generate test cases for filtering out invalid translations. These methods are developed by assuming that the validators are reliable and can be reduced to the \method{MaxPass} strategy in our work. However, it may be ineffective when the validators are plausible, as evidenced in \cref{sec:theorem}. In code generation, several cluster-based strategies are proposed to leverage incomplete or plausible test cases to rerank LLM-generated code solutions \cite{li2022competition,shi2022natural,chen2023codet}.
Li \etal~\cite{li2022competition}, Shi \etal~\cite{shi2022natural} and Chen \etal~\cite{chen2023codet} clustered code solutions based on their test results and scored each with the cluster capacity.
These cluster-based heuristics, particularly \method{CodeT}~\cite{chen2023codet}, can work well when the test cases are plausible but are susceptible to the incorrectness of solutions as in \cref{sec:theorem}. 

Some research uses deep learning techniques for ranking LLM-generated code snippets without executable test cases. Inala \etal~\cite{NEURIPS2022_5762c579} introduced a neural ranker for predicting the validity of a sampled program. Chen \etal~\cite{chen2021evaluating} and Zhang \etal~\cite{pmlr-v202-zhang23av} leveraged the LLM likelihood of the generated program for selecting the most probable code snippets. These strategies fall beyond the scope of this work since the problem we tackle does not assume the existence of additional training data or the ranking scores produced by the generation techniques. However, it is an interesting question whether these strategies have a theoretical guarantee.

\stitle{Code generation.} Code generation is an important task in software engineering, aimed at automating the production of code from defined software requirements \cite{liu2022deep}. Traditional techniques rely on predefined rules, templates, or configuration data to automate the process \cite{halbwachs_generating_1991,codegenerationforstate-based}, and often struggle with flexibility across different projects. Due to the impressive success of large language models (LLMs), recent studies focus on training LLMs on extensive code corpora to tackle complex code generation challenges \cite{zan2023large}.
Many code LLMs have shown remarkable capabilities in this domain, such as Codex \cite{chen2021evaluating}, CodeGen \cite{nijkamp2023codegen}, StarCoder \cite{li2023starcoder}, CodeLlama \cite{rozière2024code} and DeepSeek-Coder \cite{guo2024deepseek}.
This paper focuses on assessing the code solutions generated by a code generation approach with plausible test cases, and is thus orthogonal to these techniques.

\stitle{Test case generation.} Developing and maintaining human-crafted test cases can be expensive. Many techniques have been proposed to automatically generate test cases. Traditional approaches include search-based \cite{5342440,10172800,lukasczyk2022pynguin}, constrained-based \cite{6693084}, and probability-based techniques \cite{4222570}. Although most of these approaches achieve satisfactory correctness, they are constrained by inadequate coverage and poor readability, and are typically limited to generating only regression oracles \cite{xie2006augmenting} or implicit oracles \cite{barr2014oracle}. Recently, applying deep learning models (\eg, LLMs) to generate test cases has become popular \cite{alagarsamy2023a3test, tufano2021unit, Tufano_2022, 10298372, 9797060, 9401982,nie2023learning,chen2024chatunitest,dakhel2023effective,yuan2024manual,10329992,10172590}. However, ensuring the correctness and reliability of these generated test cases remains difficult. This paper explores the challenging problem of employing such plausible test cases for selecting plausible code solutions.

\section{Conclusion and future work} 

In this study, we introduce a systematic framework to derive an optimal strategy for assessing and selecting plausible code solutions using plausible test cases. We then develop a novel approach that approximates this optimal strategy with an error bound and tailors it for code generation tasks. By theoretical analysis, we show that existing heuristics are suboptimal. Our strategy substantially outperforms existing heuristics in several real-world benchmarks.

Future work could explore adapting our framework to other generation tasks in software engineering, such as automatic program repair and code translation. Also, the effectiveness of our proposed priors in these contexts, as well as the potential for alternative priors, remains an open question.

\added{\textbf{Our online appendix is available on Zenodo \cite{zenodo}}}.

\begin{acks}
This research is supported by the National Natural Science Foundation of China (No. 62202420) and the Software Engineering Application Technology Lab at Huawei under the Contract
TC20231108060. Zhongxin Liu gratefully acknowledges the support of Zhejiang University Education Foundation Qizhen Scholar Foundation. \added{We would also like to thank Yihua Sun for inspiring the incorporation of prior knowledge and for proofreading the manuscript, as well as Zinan Zhao and Junlin Chen for their discussions on the theory.}
\end{acks}

\balance
\bibliographystyle{ACM-Reference-Format}
\bibliography{reference}


\begin{thebibliography}{49}


\ifx \showCODEN    \undefined \def \showCODEN     #1{\unskip}     \fi
\ifx \showDOI      \undefined \def \showDOI       #1{#1}\fi
\ifx \showISBNx    \undefined \def \showISBNx     #1{\unskip}     \fi
\ifx \showISBNxiii \undefined \def \showISBNxiii  #1{\unskip}     \fi
\ifx \showISSN     \undefined \def \showISSN      #1{\unskip}     \fi
\ifx \showLCCN     \undefined \def \showLCCN      #1{\unskip}     \fi
\ifx \shownote     \undefined \def \shownote      #1{#1}          \fi
\ifx \showarticletitle \undefined \def \showarticletitle #1{#1}   \fi
\ifx \showURL      \undefined \def \showURL       {\relax}        \fi
\providecommand\bibfield[2]{#2}
\providecommand\bibinfo[2]{#2}
\providecommand\natexlab[1]{#1}
\providecommand\showeprint[2][]{arXiv:#2}

\bibitem[Alagarsamy et~al\mbox{.}(2023)]%
        {alagarsamy2023a3test}
\bibfield{author}{\bibinfo{person}{Saranya Alagarsamy}, \bibinfo{person}{Chakkrit Tantithamthavorn}, {and} \bibinfo{person}{Aldeida Aleti}.} \bibinfo{year}{2023}\natexlab{}.
\newblock \bibinfo{title}{A3Test: Assertion-Augmented Automated Test Case Generation}.
\newblock
\newblock
\showeprint[arxiv]{2302.10352}~[cs.SE]


\bibitem[Austin et~al\mbox{.}(2021)]%
        {austin2021program}
\bibfield{author}{\bibinfo{person}{Jacob Austin}, \bibinfo{person}{Augustus Odena}, \bibinfo{person}{Maxwell Nye}, \bibinfo{person}{Maarten Bosma}, \bibinfo{person}{Henryk Michalewski}, \bibinfo{person}{David Dohan}, \bibinfo{person}{Ellen Jiang}, \bibinfo{person}{Carrie Cai}, \bibinfo{person}{Michael Terry}, \bibinfo{person}{Quoc Le}, {and} \bibinfo{person}{Charles Sutton}.} \bibinfo{year}{2021}\natexlab{}.
\newblock \bibinfo{title}{Program Synthesis with Large Language Models}.
\newblock
\newblock
\showeprint[arxiv]{2108.07732}~[cs.PL]


\bibitem[Barr et~al\mbox{.}(2014)]%
        {barr2014oracle}
\bibfield{author}{\bibinfo{person}{Earl~T Barr}, \bibinfo{person}{Mark Harman}, \bibinfo{person}{Phil McMinn}, \bibinfo{person}{Muzammil Shahbaz}, {and} \bibinfo{person}{Shin Yoo}.} \bibinfo{year}{2014}\natexlab{}.
\newblock \showarticletitle{The oracle problem in software testing: A survey}.
\newblock \bibinfo{journal}{\emph{IEEE transactions on software engineering}} \bibinfo{volume}{41}, \bibinfo{number}{5} (\bibinfo{year}{2014}), \bibinfo{pages}{507--525}.
\newblock


\bibitem[Bayes(1763)]%
        {bayes1763lii}
\bibfield{author}{\bibinfo{person}{Thomas Bayes}.} \bibinfo{year}{1763}\natexlab{}.
\newblock \showarticletitle{LII. An essay towards solving a problem in the doctrine of chances. By the late Rev. Mr. Bayes, FRS communicated by Mr. Price, in a letter to John Canton, AMFR S}.
\newblock \bibinfo{journal}{\emph{Philosophical transactions of the Royal Society of London}} \bibinfo{number}{53} (\bibinfo{year}{1763}), \bibinfo{pages}{370--418}.
\newblock


\bibitem[Chen et~al\mbox{.}(2023)]%
        {chen2023codet}
\bibfield{author}{\bibinfo{person}{Bei Chen}, \bibinfo{person}{Fengji Zhang}, \bibinfo{person}{Anh Nguyen}, \bibinfo{person}{Daoguang Zan}, \bibinfo{person}{Zeqi Lin}, \bibinfo{person}{Jian{-}Guang Lou}, {and} \bibinfo{person}{Weizhu Chen}.} \bibinfo{year}{2023}\natexlab{}.
\newblock \showarticletitle{CodeT: Code Generation with Generated Tests}. In \bibinfo{booktitle}{\emph{The Eleventh International Conference on Learning Representations, {ICLR} 2023, Kigali, Rwanda, May 1-5, 2023}}. \bibinfo{publisher}{OpenReview.net}.
\newblock
\urldef\tempurl%
\url{https://openreview.net/pdf?id=ktrw68Cmu9c}
\showURL{%
\tempurl}


\bibitem[Chen et~al\mbox{.}(2024b)]%
        {zenodo}
\bibfield{author}{\bibinfo{person}{Mouxiang Chen}, \bibinfo{person}{Zhongxin Liu}, \bibinfo{person}{He Tao}, \bibinfo{person}{Yusu Hong}, \bibinfo{person}{David Lo}, \bibinfo{person}{Xin Xia}, {and} \bibinfo{person}{Jianling Sun}.} \bibinfo{year}{2024}\natexlab{b}.
\newblock \bibinfo{booktitle}{\emph{{B4: Towards Optimal Assessment of Plausible Code Solutions with Plausible Tests}}}.
\newblock
\urldef\tempurl%
\url{https://doi.org/10.5281/zenodo.13737381}
\showDOI{\tempurl}


\bibitem[Chen et~al\mbox{.}(2021)]%
        {chen2021evaluating}
\bibfield{author}{\bibinfo{person}{Mark Chen}, \bibinfo{person}{Jerry Tworek}, \bibinfo{person}{Heewoo Jun}, \bibinfo{person}{Qiming Yuan}, \bibinfo{person}{Henrique~Ponde de Oliveira~Pinto}, \bibinfo{person}{Jared Kaplan}, \bibinfo{person}{Harri Edwards}, \bibinfo{person}{Yuri Burda}, \bibinfo{person}{Nicholas Joseph}, \bibinfo{person}{Greg Brockman}, \bibinfo{person}{Alex Ray}, \bibinfo{person}{Raul Puri}, \bibinfo{person}{Gretchen Krueger}, \bibinfo{person}{Michael Petrov}, \bibinfo{person}{Heidy Khlaaf}, \bibinfo{person}{Girish Sastry}, \bibinfo{person}{Pamela Mishkin}, \bibinfo{person}{Brooke Chan}, \bibinfo{person}{Scott Gray}, \bibinfo{person}{Nick Ryder}, \bibinfo{person}{Mikhail Pavlov}, \bibinfo{person}{Alethea Power}, \bibinfo{person}{Lukasz Kaiser}, \bibinfo{person}{Mohammad Bavarian}, \bibinfo{person}{Clemens Winter}, \bibinfo{person}{Philippe Tillet}, \bibinfo{person}{Felipe~Petroski Such}, \bibinfo{person}{Dave Cummings}, \bibinfo{person}{Matthias Plappert}, \bibinfo{person}{Fotios Chantzis},
  \bibinfo{person}{Elizabeth Barnes}, \bibinfo{person}{Ariel Herbert-Voss}, \bibinfo{person}{William~Hebgen Guss}, \bibinfo{person}{Alex Nichol}, \bibinfo{person}{Alex Paino}, \bibinfo{person}{Nikolas Tezak}, \bibinfo{person}{Jie Tang}, \bibinfo{person}{Igor Babuschkin}, \bibinfo{person}{Suchir Balaji}, \bibinfo{person}{Shantanu Jain}, \bibinfo{person}{William Saunders}, \bibinfo{person}{Christopher Hesse}, \bibinfo{person}{Andrew~N. Carr}, \bibinfo{person}{Jan Leike}, \bibinfo{person}{Josh Achiam}, \bibinfo{person}{Vedant Misra}, \bibinfo{person}{Evan Morikawa}, \bibinfo{person}{Alec Radford}, \bibinfo{person}{Matthew Knight}, \bibinfo{person}{Miles Brundage}, \bibinfo{person}{Mira Murati}, \bibinfo{person}{Katie Mayer}, \bibinfo{person}{Peter Welinder}, \bibinfo{person}{Bob McGrew}, \bibinfo{person}{Dario Amodei}, \bibinfo{person}{Sam McCandlish}, \bibinfo{person}{Ilya Sutskever}, {and} \bibinfo{person}{Wojciech Zaremba}.} \bibinfo{year}{2021}\natexlab{}.
\newblock \bibinfo{title}{Evaluating Large Language Models Trained on Code}.
\newblock
\newblock
\showeprint[arxiv]{2107.03374}~[cs.LG]


\bibitem[Chen et~al\mbox{.}(2024a)]%
        {chen2024chatunitest}
\bibfield{author}{\bibinfo{person}{Yinghao Chen}, \bibinfo{person}{Zehao Hu}, \bibinfo{person}{Chen Zhi}, \bibinfo{person}{Junxiao Han}, \bibinfo{person}{Shuiguang Deng}, {and} \bibinfo{person}{Jianwei Yin}.} \bibinfo{year}{2024}\natexlab{a}.
\newblock \bibinfo{title}{ChatUniTest: A Framework for LLM-Based Test Generation}.
\newblock
\newblock
\showeprint[arxiv]{2305.04764}~[cs.SE]


\bibitem[Dakhel et~al\mbox{.}(2023)]%
        {dakhel2023effective}
\bibfield{author}{\bibinfo{person}{Arghavan~Moradi Dakhel}, \bibinfo{person}{Amin Nikanjam}, \bibinfo{person}{Vahid Majdinasab}, \bibinfo{person}{Foutse Khomh}, {and} \bibinfo{person}{Michel~C. Desmarais}.} \bibinfo{year}{2023}\natexlab{}.
\newblock \bibinfo{title}{Effective Test Generation Using Pre-trained Large Language Models and Mutation Testing}.
\newblock
\newblock
\showeprint[arxiv]{2308.16557}~[cs.SE]


\bibitem[Davis(1972)]%
        {davis1972gamma}
\bibfield{author}{\bibinfo{person}{Philip~J Davis}.} \bibinfo{year}{1972}\natexlab{}.
\newblock \showarticletitle{Gamma function and related functions}.
\newblock \bibinfo{journal}{\emph{Handbook of mathematical functions}}  \bibinfo{volume}{256} (\bibinfo{year}{1972}).
\newblock


\bibitem[DeGroot(2005)]%
        {degroot2005optimal}
\bibfield{author}{\bibinfo{person}{Morris~H DeGroot}.} \bibinfo{year}{2005}\natexlab{}.
\newblock \bibinfo{booktitle}{\emph{Optimal statistical decisions}}.
\newblock \bibinfo{publisher}{John Wiley \& Sons}.
\newblock


\bibitem[Fraser and Arcuri(2011)]%
        {fraser2011evosuite}
\bibfield{author}{\bibinfo{person}{Gordon Fraser} {and} \bibinfo{person}{Andrea Arcuri}.} \bibinfo{year}{2011}\natexlab{}.
\newblock \showarticletitle{Evosuite: automatic test suite generation for object-oriented software}. In \bibinfo{booktitle}{\emph{Proceedings of the 19th ACM SIGSOFT symposium and the 13th European conference on Foundations of software engineering}}. \bibinfo{pages}{416--419}.
\newblock


\bibitem[Guo et~al\mbox{.}(2024)]%
        {guo2024deepseek}
\bibfield{author}{\bibinfo{person}{Daya Guo}, \bibinfo{person}{Qihao Zhu}, \bibinfo{person}{Dejian Yang}, \bibinfo{person}{Zhenda Xie}, \bibinfo{person}{Kai Dong}, \bibinfo{person}{Wentao Zhang}, \bibinfo{person}{Guanting Chen}, \bibinfo{person}{Xiao Bi}, \bibinfo{person}{Y Wu}, \bibinfo{person}{YK Li}, {et~al\mbox{.}}} \bibinfo{year}{2024}\natexlab{}.
\newblock \showarticletitle{DeepSeek-Coder: When the Large Language Model Meets Programming--The Rise of Code Intelligence}.
\newblock \bibinfo{journal}{\emph{arXiv preprint arXiv:2401.14196}} (\bibinfo{year}{2024}).
\newblock


\bibitem[Halbwachs et~al\mbox{.}(1991)]%
        {halbwachs_generating_1991}
\bibfield{author}{\bibinfo{person}{Nicolas Halbwachs}, \bibinfo{person}{Pascal Raymond}, {and} \bibinfo{person}{Christophe Ratel}.} \bibinfo{year}{1991}\natexlab{}.
\newblock \showarticletitle{Generating efficient code from data-flow programs}. In \bibinfo{booktitle}{\emph{Programming Language Implementation and Logic Programming: 3rd International Symposium, PLILP'91 Passau, Germany, August 26--28, 1991 Proceedings 3}}. Springer, \bibinfo{pages}{207--218}.
\newblock


\bibitem[Harman and McMinn(2010)]%
        {5342440}
\bibfield{author}{\bibinfo{person}{Mark Harman} {and} \bibinfo{person}{Phil McMinn}.} \bibinfo{year}{2010}\natexlab{}.
\newblock \showarticletitle{A Theoretical and Empirical Study of Search-Based Testing: Local, Global, and Hybrid Search}.
\newblock \bibinfo{journal}{\emph{IEEE Transactions on Software Engineering}} \bibinfo{volume}{36}, \bibinfo{number}{2} (\bibinfo{year}{2010}), \bibinfo{pages}{226--247}.
\newblock
\urldef\tempurl%
\url{https://doi.org/10.1109/TSE.2009.71}
\showDOI{\tempurl}


\bibitem[Hendrycks et~al\mbox{.}(2021)]%
        {hendrycks2021measuring}
\bibfield{author}{\bibinfo{person}{Dan Hendrycks}, \bibinfo{person}{Steven Basart}, \bibinfo{person}{Saurav Kadavath}, \bibinfo{person}{Mantas Mazeika}, \bibinfo{person}{Akul Arora}, \bibinfo{person}{Ethan Guo}, \bibinfo{person}{Collin Burns}, \bibinfo{person}{Samir Puranik}, \bibinfo{person}{Horace He}, \bibinfo{person}{Dawn Song}, {and} \bibinfo{person}{Jacob Steinhardt}.} \bibinfo{year}{2021}\natexlab{}.
\newblock \showarticletitle{Measuring Coding Challenge Competence With {APPS}}. In \bibinfo{booktitle}{\emph{Thirty-fifth Conference on Neural Information Processing Systems Datasets and Benchmarks Track (Round 2)}}.
\newblock
\urldef\tempurl%
\url{https://openreview.net/forum?id=sD93GOzH3i5}
\showURL{%
\tempurl}


\bibitem[Inala et~al\mbox{.}(2022)]%
        {NEURIPS2022_5762c579}
\bibfield{author}{\bibinfo{person}{Jeevana~Priya Inala}, \bibinfo{person}{Chenglong Wang}, \bibinfo{person}{Mei Yang}, \bibinfo{person}{Andres Codas}, \bibinfo{person}{Mark Encarnaci\'{o}n}, \bibinfo{person}{Shuvendu Lahiri}, \bibinfo{person}{Madanlal Musuvathi}, {and} \bibinfo{person}{Jianfeng Gao}.} \bibinfo{year}{2022}\natexlab{}.
\newblock \showarticletitle{Fault-Aware Neural Code Rankers}. In \bibinfo{booktitle}{\emph{Advances in Neural Information Processing Systems}}, \bibfield{editor}{\bibinfo{person}{S.~Koyejo}, \bibinfo{person}{S.~Mohamed}, \bibinfo{person}{A.~Agarwal}, \bibinfo{person}{D.~Belgrave}, \bibinfo{person}{K.~Cho}, {and} \bibinfo{person}{A.~Oh}} (Eds.), Vol.~\bibinfo{volume}{35}. \bibinfo{publisher}{Curran Associates, Inc.}, \bibinfo{pages}{13419--13432}.
\newblock
\urldef\tempurl%
\url{https://proceedings.neurips.cc/paper_files/paper/2022/file/5762c579d09811b7639be2389b3d07be-Paper-Conference.pdf}
\showURL{%
\tempurl}


\bibitem[Lahiri et~al\mbox{.}(2023)]%
        {lahiri2023interactive}
\bibfield{author}{\bibinfo{person}{Shuvendu~K. Lahiri}, \bibinfo{person}{Sarah Fakhoury}, \bibinfo{person}{Aaditya Naik}, \bibinfo{person}{Georgios Sakkas}, \bibinfo{person}{Saikat Chakraborty}, \bibinfo{person}{Madanlal Musuvathi}, \bibinfo{person}{Piali Choudhury}, \bibinfo{person}{Curtis von Veh}, \bibinfo{person}{Jeevana~Priya Inala}, \bibinfo{person}{Chenglong Wang}, {and} \bibinfo{person}{Jianfeng Gao}.} \bibinfo{year}{2023}\natexlab{}.
\newblock \bibinfo{title}{Interactive Code Generation via Test-Driven User-Intent Formalization}.
\newblock
\newblock
\showeprint[arxiv]{2208.05950}~[cs.SE]


\bibitem[Le et~al\mbox{.}(2022)]%
        {le2022coderl}
\bibfield{author}{\bibinfo{person}{Hung Le}, \bibinfo{person}{Yue Wang}, \bibinfo{person}{Akhilesh~Deepak Gotmare}, \bibinfo{person}{Silvio Savarese}, {and} \bibinfo{person}{Steven Chu~Hong Hoi}.} \bibinfo{year}{2022}\natexlab{}.
\newblock \showarticletitle{Coderl: Mastering code generation through pretrained models and deep reinforcement learning}.
\newblock \bibinfo{journal}{\emph{Advances in Neural Information Processing Systems}}  \bibinfo{volume}{35} (\bibinfo{year}{2022}), \bibinfo{pages}{21314--21328}.
\newblock


\bibitem[Lemieux et~al\mbox{.}(2023)]%
        {10172800}
\bibfield{author}{\bibinfo{person}{Caroline Lemieux}, \bibinfo{person}{Jeevana~Priya Inala}, \bibinfo{person}{Shuvendu~K. Lahiri}, {and} \bibinfo{person}{Siddhartha Sen}.} \bibinfo{year}{2023}\natexlab{}.
\newblock \showarticletitle{CodaMosa: Escaping Coverage Plateaus in Test Generation with Pre-trained Large Language Models}. In \bibinfo{booktitle}{\emph{2023 IEEE/ACM 45th International Conference on Software Engineering (ICSE)}}. \bibinfo{pages}{919--931}.
\newblock
\urldef\tempurl%
\url{https://doi.org/10.1109/ICSE48619.2023.00085}
\showDOI{\tempurl}


\bibitem[Li et~al\mbox{.}(2023)]%
        {li2023starcoder}
\bibfield{author}{\bibinfo{person}{Raymond Li}, \bibinfo{person}{Loubna~Ben allal}, \bibinfo{person}{Yangtian Zi}, \bibinfo{person}{Niklas Muennighoff}, \bibinfo{person}{Denis Kocetkov}, \bibinfo{person}{Chenghao Mou}, \bibinfo{person}{Marc Marone}, \bibinfo{person}{Christopher Akiki}, \bibinfo{person}{Jia LI}, \bibinfo{person}{Jenny Chim}, \bibinfo{person}{Qian Liu}, \bibinfo{person}{Evgenii Zheltonozhskii}, \bibinfo{person}{Terry~Yue Zhuo}, \bibinfo{person}{Thomas Wang}, \bibinfo{person}{Olivier Dehaene}, \bibinfo{person}{Joel Lamy-Poirier}, \bibinfo{person}{Joao Monteiro}, \bibinfo{person}{Nicolas Gontier}, \bibinfo{person}{Ming-Ho Yee}, \bibinfo{person}{Logesh~Kumar Umapathi}, \bibinfo{person}{Jian Zhu}, \bibinfo{person}{Ben Lipkin}, \bibinfo{person}{Muhtasham Oblokulov}, \bibinfo{person}{Zhiruo Wang}, \bibinfo{person}{Rudra Murthy}, \bibinfo{person}{Jason~T Stillerman}, \bibinfo{person}{Siva~Sankalp Patel}, \bibinfo{person}{Dmitry Abulkhanov}, \bibinfo{person}{Marco Zocca}, \bibinfo{person}{Manan Dey},
  \bibinfo{person}{Zhihan Zhang}, \bibinfo{person}{Urvashi Bhattacharyya}, \bibinfo{person}{Wenhao Yu}, \bibinfo{person}{Sasha Luccioni}, \bibinfo{person}{Paulo Villegas}, \bibinfo{person}{Fedor Zhdanov}, \bibinfo{person}{Tony Lee}, \bibinfo{person}{Nadav Timor}, \bibinfo{person}{Jennifer Ding}, \bibinfo{person}{Claire~S Schlesinger}, \bibinfo{person}{Hailey Schoelkopf}, \bibinfo{person}{Jan Ebert}, \bibinfo{person}{Tri Dao}, \bibinfo{person}{Mayank Mishra}, \bibinfo{person}{Alex Gu}, \bibinfo{person}{Carolyn~Jane Anderson}, \bibinfo{person}{Brendan Dolan-Gavitt}, \bibinfo{person}{Danish Contractor}, \bibinfo{person}{Siva Reddy}, \bibinfo{person}{Daniel Fried}, \bibinfo{person}{Dzmitry Bahdanau}, \bibinfo{person}{Yacine Jernite}, \bibinfo{person}{Carlos~Mu{\~n}oz Ferrandis}, \bibinfo{person}{Sean Hughes}, \bibinfo{person}{Thomas Wolf}, \bibinfo{person}{Arjun Guha}, \bibinfo{person}{Leandro~Von Werra}, {and} \bibinfo{person}{Harm de Vries}.} \bibinfo{year}{2023}\natexlab{}.
\newblock \showarticletitle{StarCoder: may the source be with you!}
\newblock \bibinfo{journal}{\emph{Transactions on Machine Learning Research}} (\bibinfo{year}{2023}).
\newblock
\showISSN{2835-8856}
\urldef\tempurl%
\url{https://openreview.net/forum?id=KoFOg41haE}
\showURL{%
\tempurl}
\newblock
\shownote{Reproducibility Certification}.


\bibitem[Li et~al\mbox{.}(2022)]%
        {li2022competition}
\bibfield{author}{\bibinfo{person}{Yujia Li}, \bibinfo{person}{David Choi}, \bibinfo{person}{Junyoung Chung}, \bibinfo{person}{Nate Kushman}, \bibinfo{person}{Julian Schrittwieser}, \bibinfo{person}{Rémi Leblond}, \bibinfo{person}{Tom Eccles}, \bibinfo{person}{James Keeling}, \bibinfo{person}{Felix Gimeno}, \bibinfo{person}{Agustin~Dal Lago}, \bibinfo{person}{Thomas Hubert}, \bibinfo{person}{Peter Choy}, \bibinfo{person}{Cyprien de Masson~d’Autume}, \bibinfo{person}{Igor Babuschkin}, \bibinfo{person}{Xinyun Chen}, \bibinfo{person}{Po-Sen Huang}, \bibinfo{person}{Johannes Welbl}, \bibinfo{person}{Sven Gowal}, \bibinfo{person}{Alexey Cherepanov}, \bibinfo{person}{James Molloy}, \bibinfo{person}{Daniel~J. Mankowitz}, \bibinfo{person}{Esme~Sutherland Robson}, \bibinfo{person}{Pushmeet Kohli}, \bibinfo{person}{Nando de Freitas}, \bibinfo{person}{Koray Kavukcuoglu}, {and} \bibinfo{person}{Oriol Vinyals}.} \bibinfo{year}{2022}\natexlab{}.
\newblock \showarticletitle{Competition-level code generation with AlphaCode}.
\newblock \bibinfo{journal}{\emph{Science}} \bibinfo{volume}{378}, \bibinfo{number}{6624} (\bibinfo{year}{2022}), \bibinfo{pages}{1092--1097}.
\newblock
\urldef\tempurl%
\url{https://doi.org/10.1126/science.abq1158}
\showDOI{\tempurl}
\showeprint{https://www.science.org/doi/pdf/10.1126/science.abq1158}


\bibitem[Liu et~al\mbox{.}(2022)]%
        {liu2022deep}
\bibfield{author}{\bibinfo{person}{Hui Liu}, \bibinfo{person}{Mingzhu Shen}, \bibinfo{person}{Jiaqi Zhu}, \bibinfo{person}{Nan Niu}, \bibinfo{person}{Ge Li}, {and} \bibinfo{person}{Lu Zhang}.} \bibinfo{year}{2022}\natexlab{}.
\newblock \showarticletitle{Deep Learning Based Program Generation From Requirements Text: Are We There Yet?}
\newblock \bibinfo{journal}{\emph{IEEE Transactions on Software Engineering}} \bibinfo{volume}{48}, \bibinfo{number}{4} (\bibinfo{year}{2022}), \bibinfo{pages}{1268--1289}.
\newblock
\urldef\tempurl%
\url{https://doi.org/10.1109/TSE.2020.3018481}
\showDOI{\tempurl}


\bibitem[Lukasczyk and Fraser(2022)]%
        {lukasczyk2022pynguin}
\bibfield{author}{\bibinfo{person}{Stephan Lukasczyk} {and} \bibinfo{person}{Gordon Fraser}.} \bibinfo{year}{2022}\natexlab{}.
\newblock \showarticletitle{Pynguin: Automated unit test generation for python}. In \bibinfo{booktitle}{\emph{Proceedings of the ACM/IEEE 44th International Conference on Software Engineering: Companion Proceedings}}. \bibinfo{pages}{168--172}.
\newblock


\bibitem[Mastropaolo et~al\mbox{.}(2023)]%
        {9797060}
\bibfield{author}{\bibinfo{person}{Antonio Mastropaolo}, \bibinfo{person}{Nathan Cooper}, \bibinfo{person}{David~Nader Palacio}, \bibinfo{person}{Simone Scalabrino}, \bibinfo{person}{Denys Poshyvanyk}, \bibinfo{person}{Rocco Oliveto}, {and} \bibinfo{person}{Gabriele Bavota}.} \bibinfo{year}{2023}\natexlab{}.
\newblock \showarticletitle{Using Transfer Learning for Code-Related Tasks}.
\newblock \bibinfo{journal}{\emph{IEEE Transactions on Software Engineering}} \bibinfo{volume}{49}, \bibinfo{number}{4} (\bibinfo{year}{2023}), \bibinfo{pages}{1580--1598}.
\newblock
\urldef\tempurl%
\url{https://doi.org/10.1109/TSE.2022.3183297}
\showDOI{\tempurl}


\bibitem[Mastropaolo et~al\mbox{.}(2021)]%
        {9401982}
\bibfield{author}{\bibinfo{person}{Antonio Mastropaolo}, \bibinfo{person}{Simone Scalabrino}, \bibinfo{person}{Nathan Cooper}, \bibinfo{person}{David Nader~Palacio}, \bibinfo{person}{Denys Poshyvanyk}, \bibinfo{person}{Rocco Oliveto}, {and} \bibinfo{person}{Gabriele Bavota}.} \bibinfo{year}{2021}\natexlab{}.
\newblock \showarticletitle{Studying the Usage of Text-To-Text Transfer Transformer to Support Code-Related Tasks}. In \bibinfo{booktitle}{\emph{2021 IEEE/ACM 43rd International Conference on Software Engineering (ICSE)}}. \bibinfo{pages}{336--347}.
\newblock
\urldef\tempurl%
\url{https://doi.org/10.1109/ICSE43902.2021.00041}
\showDOI{\tempurl}


\bibitem[Nashid et~al\mbox{.}(2023)]%
        {10172590}
\bibfield{author}{\bibinfo{person}{Noor Nashid}, \bibinfo{person}{Mifta Sintaha}, {and} \bibinfo{person}{Ali Mesbah}.} \bibinfo{year}{2023}\natexlab{}.
\newblock \showarticletitle{Retrieval-Based Prompt Selection for Code-Related Few-Shot Learning}. In \bibinfo{booktitle}{\emph{2023 IEEE/ACM 45th International Conference on Software Engineering (ICSE)}}. \bibinfo{pages}{2450--2462}.
\newblock
\urldef\tempurl%
\url{https://doi.org/10.1109/ICSE48619.2023.00205}
\showDOI{\tempurl}


\bibitem[Nie et~al\mbox{.}(2023)]%
        {nie2023learning}
\bibfield{author}{\bibinfo{person}{Pengyu Nie}, \bibinfo{person}{Rahul Banerjee}, \bibinfo{person}{Junyi~Jessy Li}, \bibinfo{person}{Raymond~J Mooney}, {and} \bibinfo{person}{Milos Gligoric}.} \bibinfo{year}{2023}\natexlab{}.
\newblock \showarticletitle{Learning deep semantics for test completion}. In \bibinfo{booktitle}{\emph{2023 IEEE/ACM 45th International Conference on Software Engineering (ICSE)}}. IEEE, \bibinfo{pages}{2111--2123}.
\newblock


\bibitem[Nijkamp et~al\mbox{.}(2023)]%
        {nijkamp2023codegen}
\bibfield{author}{\bibinfo{person}{Erik Nijkamp}, \bibinfo{person}{Bo Pang}, \bibinfo{person}{Hiroaki Hayashi}, \bibinfo{person}{Lifu Tu}, \bibinfo{person}{Huan Wang}, \bibinfo{person}{Yingbo Zhou}, \bibinfo{person}{Silvio Savarese}, {and} \bibinfo{person}{Caiming Xiong}.} \bibinfo{year}{2023}\natexlab{}.
\newblock \showarticletitle{CodeGen: An Open Large Language Model for Code with Multi-Turn Program Synthesis}. In \bibinfo{booktitle}{\emph{The Eleventh International Conference on Learning Representations}}.
\newblock
\urldef\tempurl%
\url{https://openreview.net/forum?id=iaYcJKpY2B_}
\showURL{%
\tempurl}


\bibitem[Pacheco et~al\mbox{.}(2007)]%
        {4222570}
\bibfield{author}{\bibinfo{person}{Carlos Pacheco}, \bibinfo{person}{Shuvendu~K. Lahiri}, \bibinfo{person}{Michael~D. Ernst}, {and} \bibinfo{person}{Thomas Ball}.} \bibinfo{year}{2007}\natexlab{}.
\newblock \showarticletitle{Feedback-Directed Random Test Generation}. In \bibinfo{booktitle}{\emph{29th International Conference on Software Engineering (ICSE'07)}}. \bibinfo{pages}{75--84}.
\newblock
\urldef\tempurl%
\url{https://doi.org/10.1109/ICSE.2007.37}
\showDOI{\tempurl}


\bibitem[Raiffa and Schlaifer(2000)]%
        {raiffa2000applied}
\bibfield{author}{\bibinfo{person}{Howard Raiffa} {and} \bibinfo{person}{Robert Schlaifer}.} \bibinfo{year}{2000}\natexlab{}.
\newblock \bibinfo{booktitle}{\emph{Applied statistical decision theory}}. Vol.~\bibinfo{volume}{78}.
\newblock \bibinfo{publisher}{John Wiley \& Sons}.
\newblock


\bibitem[Rao et~al\mbox{.}(2023)]%
        {10298372}
\bibfield{author}{\bibinfo{person}{Nikitha Rao}, \bibinfo{person}{Kush Jain}, \bibinfo{person}{Uri Alon}, \bibinfo{person}{Claire~Le Goues}, {and} \bibinfo{person}{Vincent~J. Hellendoorn}.} \bibinfo{year}{2023}\natexlab{}.
\newblock \showarticletitle{CAT-LM Training Language Models on Aligned Code And Tests}. In \bibinfo{booktitle}{\emph{2023 38th IEEE/ACM International Conference on Automated Software Engineering (ASE)}}. \bibinfo{pages}{409--420}.
\newblock
\urldef\tempurl%
\url{https://doi.org/10.1109/ASE56229.2023.00193}
\showDOI{\tempurl}


\bibitem[Roziere et~al\mbox{.}(2022)]%
        {roziere2022leveraging}
\bibfield{author}{\bibinfo{person}{Baptiste Roziere}, \bibinfo{person}{Jie Zhang}, \bibinfo{person}{Francois Charton}, \bibinfo{person}{Mark Harman}, \bibinfo{person}{Gabriel Synnaeve}, {and} \bibinfo{person}{Guillaume Lample}.} \bibinfo{year}{2022}\natexlab{}.
\newblock \showarticletitle{Leveraging Automated Unit Tests for Unsupervised Code Translation}. In \bibinfo{booktitle}{\emph{International Conference on Learning Representations}}.
\newblock
\urldef\tempurl%
\url{https://openreview.net/forum?id=cmt-6KtR4c4}
\showURL{%
\tempurl}


\bibitem[Rozière et~al\mbox{.}(2024)]%
        {rozière2024code}
\bibfield{author}{\bibinfo{person}{Baptiste Rozière}, \bibinfo{person}{Jonas Gehring}, \bibinfo{person}{Fabian Gloeckle}, \bibinfo{person}{Sten Sootla}, \bibinfo{person}{Itai Gat}, \bibinfo{person}{Xiaoqing~Ellen Tan}, \bibinfo{person}{Yossi Adi}, \bibinfo{person}{Jingyu Liu}, \bibinfo{person}{Romain Sauvestre}, \bibinfo{person}{Tal Remez}, \bibinfo{person}{Jérémy Rapin}, \bibinfo{person}{Artyom Kozhevnikov}, \bibinfo{person}{Ivan Evtimov}, \bibinfo{person}{Joanna Bitton}, \bibinfo{person}{Manish Bhatt}, \bibinfo{person}{Cristian~Canton Ferrer}, \bibinfo{person}{Aaron Grattafiori}, \bibinfo{person}{Wenhan Xiong}, \bibinfo{person}{Alexandre Défossez}, \bibinfo{person}{Jade Copet}, \bibinfo{person}{Faisal Azhar}, \bibinfo{person}{Hugo Touvron}, \bibinfo{person}{Louis Martin}, \bibinfo{person}{Nicolas Usunier}, \bibinfo{person}{Thomas Scialom}, {and} \bibinfo{person}{Gabriel Synnaeve}.} \bibinfo{year}{2024}\natexlab{}.
\newblock \bibinfo{title}{Code Llama: Open Foundation Models for Code}.
\newblock
\newblock
\showeprint[arxiv]{2308.12950}~[cs.CL]


\bibitem[Schäfer et~al\mbox{.}(2024)]%
        {10329992}
\bibfield{author}{\bibinfo{person}{Max Schäfer}, \bibinfo{person}{Sarah Nadi}, \bibinfo{person}{Aryaz Eghbali}, {and} \bibinfo{person}{Frank Tip}.} \bibinfo{year}{2024}\natexlab{}.
\newblock \showarticletitle{An Empirical Evaluation of Using Large Language Models for Automated Unit Test Generation}.
\newblock \bibinfo{journal}{\emph{IEEE Transactions on Software Engineering}} \bibinfo{volume}{50}, \bibinfo{number}{1} (\bibinfo{year}{2024}), \bibinfo{pages}{85--105}.
\newblock
\urldef\tempurl%
\url{https://doi.org/10.1109/TSE.2023.3334955}
\showDOI{\tempurl}


\bibitem[Shi et~al\mbox{.}(2022)]%
        {shi2022natural}
\bibfield{author}{\bibinfo{person}{Freda Shi}, \bibinfo{person}{Daniel Fried}, \bibinfo{person}{Marjan Ghazvininejad}, \bibinfo{person}{Luke Zettlemoyer}, {and} \bibinfo{person}{Sida~I Wang}.} \bibinfo{year}{2022}\natexlab{}.
\newblock \showarticletitle{Natural Language to Code Translation with Execution}. In \bibinfo{booktitle}{\emph{Proceedings of the 2022 Conference on Empirical Methods in Natural Language Processing}}. \bibinfo{pages}{3533--3546}.
\newblock


\bibitem[Stojanac et~al\mbox{.}(2017)]%
        {stojanac2017products}
\bibfield{author}{\bibinfo{person}{{\v{Z}}eljka Stojanac}, \bibinfo{person}{Daniel Suess}, {and} \bibinfo{person}{Martin Kliesch}.} \bibinfo{year}{2017}\natexlab{}.
\newblock \showarticletitle{On products of Gaussian random variables}.
\newblock \bibinfo{journal}{\emph{arXiv preprint arXiv:1711.10516}} (\bibinfo{year}{2017}).
\newblock


\bibitem[Tsybakov(2008)]%
        {tsybakov2008introduction}
\bibfield{author}{\bibinfo{person}{A.B. Tsybakov}.} \bibinfo{year}{2008}\natexlab{}.
\newblock \bibinfo{booktitle}{\emph{Introduction to Nonparametric Estimation}}.
\newblock \bibinfo{publisher}{Springer New York}.
\newblock
\showISBNx{9780387790527}
\showLCCN{2008939894}
\urldef\tempurl%
\url{https://books.google.com.hk/books?id=mwB8rUBsbqoC}
\showURL{%
\tempurl}


\bibitem[Tufano et~al\mbox{.}(2021)]%
        {tufano2021unit}
\bibfield{author}{\bibinfo{person}{Michele Tufano}, \bibinfo{person}{Dawn Drain}, \bibinfo{person}{Alexey Svyatkovskiy}, \bibinfo{person}{Shao~Kun Deng}, {and} \bibinfo{person}{Neel Sundaresan}.} \bibinfo{year}{2021}\natexlab{}.
\newblock \bibinfo{title}{Unit Test Case Generation with Transformers and Focal Context}.
\newblock
\newblock
\showeprint[arxiv]{2009.05617}~[cs.SE]


\bibitem[Tufano et~al\mbox{.}(2022)]%
        {Tufano_2022}
\bibfield{author}{\bibinfo{person}{Michele Tufano}, \bibinfo{person}{Dawn Drain}, \bibinfo{person}{Alexey Svyatkovskiy}, {and} \bibinfo{person}{Neel Sundaresan}.} \bibinfo{year}{2022}\natexlab{}.
\newblock \showarticletitle{Generating accurate assert statements for unit test cases using pretrained transformers}. In \bibinfo{booktitle}{\emph{Proceedings of the 3rd ACM/IEEE International Conference on Automation of Software Test}} \emph{(\bibinfo{series}{AST ’22})}. \bibinfo{publisher}{ACM}.
\newblock
\urldef\tempurl%
\url{https://doi.org/10.1145/3524481.3527220}
\showDOI{\tempurl}


\bibitem[Virtanen et~al\mbox{.}(2020)]%
        {pauli2020scipy}
\bibfield{author}{\bibinfo{person}{Pauli Virtanen}, \bibinfo{person}{Ralf Gommers}, \bibinfo{person}{Travis~E. Oliphant}, \bibinfo{person}{Matt Haberland}, \bibinfo{person}{Tyler Reddy}, \bibinfo{person}{David Cournapeau}, \bibinfo{person}{Evgeni Burovski}, \bibinfo{person}{Pearu Peterson}, \bibinfo{person}{Warren Weckesser}, \bibinfo{person}{Jonathan Bright}, \bibinfo{person}{St{\'e}fan~J. {van der Walt}}, \bibinfo{person}{Matthew Brett}, \bibinfo{person}{Joshua Wilson}, \bibinfo{person}{K.~Jarrod Millman}, \bibinfo{person}{Nikolay Mayorov}, \bibinfo{person}{Andrew R.~J. Nelson}, \bibinfo{person}{Eric Jones}, \bibinfo{person}{Robert Kern}, \bibinfo{person}{Eric Larson}, \bibinfo{person}{C~J Carey}, \bibinfo{person}{{\.I}lhan Polat}, \bibinfo{person}{Yu Feng}, \bibinfo{person}{Eric~W. Moore}, \bibinfo{person}{Jake {VanderPlas}}, \bibinfo{person}{Denis Laxalde}, \bibinfo{person}{Josef Perktold}, \bibinfo{person}{Robert Cimrman}, \bibinfo{person}{Ian Henriksen}, \bibinfo{person}{E.~A. Quintero},
  \bibinfo{person}{Charles~R. Harris}, \bibinfo{person}{Anne~M. Archibald}, \bibinfo{person}{Ant{\^o}nio~H. Ribeiro}, \bibinfo{person}{Fabian Pedregosa}, \bibinfo{person}{Paul {van Mulbregt}}, {and} \bibinfo{person}{{SciPy 1.0 Contributors}}.} \bibinfo{year}{2020}\natexlab{}.
\newblock \showarticletitle{{{SciPy} 1.0: Fundamental Algorithms for Scientific Computing in Python}}.
\newblock \bibinfo{journal}{\emph{Nature Methods}}  \bibinfo{volume}{17} (\bibinfo{year}{2020}), \bibinfo{pages}{261--272}.
\newblock
\urldef\tempurl%
\url{https://doi.org/10.1038/s41592-019-0686-2}
\showDOI{\tempurl}


\bibitem[Whalen(2000)]%
        {codegenerationforstate-based}
\bibfield{author}{\bibinfo{person}{Michael~W Whalen}.} \bibinfo{year}{2000}\natexlab{}.
\newblock \showarticletitle{High-integrity code generation for state-based formalisms}. In \bibinfo{booktitle}{\emph{Proceedings of the 22nd international conference on Software engineering}}. \bibinfo{pages}{725--727}.
\newblock


\bibitem[Wilcoxon(1992)]%
        {wilcoxon1992individual}
\bibfield{author}{\bibinfo{person}{Frank Wilcoxon}.} \bibinfo{year}{1992}\natexlab{}.
\newblock \showarticletitle{Individual comparisons by ranking methods}.
\newblock In \bibinfo{booktitle}{\emph{Breakthroughs in statistics: Methodology and distribution}}. \bibinfo{publisher}{Springer}, \bibinfo{pages}{196--202}.
\newblock


\bibitem[Xiao et~al\mbox{.}(2013)]%
        {6693084}
\bibfield{author}{\bibinfo{person}{Xusheng Xiao}, \bibinfo{person}{Sihan Li}, \bibinfo{person}{Tao Xie}, {and} \bibinfo{person}{Nikolai Tillmann}.} \bibinfo{year}{2013}\natexlab{}.
\newblock \showarticletitle{Characteristic studies of loop problems for structural test generation via symbolic execution}. In \bibinfo{booktitle}{\emph{2013 28th IEEE/ACM International Conference on Automated Software Engineering (ASE)}}. \bibinfo{pages}{246--256}.
\newblock
\urldef\tempurl%
\url{https://doi.org/10.1109/ASE.2013.6693084}
\showDOI{\tempurl}


\bibitem[Xie(2006)]%
        {xie2006augmenting}
\bibfield{author}{\bibinfo{person}{Tao Xie}.} \bibinfo{year}{2006}\natexlab{}.
\newblock \showarticletitle{Augmenting automatically generated unit-test suites with regression oracle checking}. In \bibinfo{booktitle}{\emph{European Conference on Object-Oriented Programming}}. Springer, \bibinfo{pages}{380--403}.
\newblock


\bibitem[Yang et~al\mbox{.}(2017)]%
        {yang2017better}
\bibfield{author}{\bibinfo{person}{Jinqiu Yang}, \bibinfo{person}{Alexey Zhikhartsev}, \bibinfo{person}{Yuefei Liu}, {and} \bibinfo{person}{Lin Tan}.} \bibinfo{year}{2017}\natexlab{}.
\newblock \showarticletitle{Better test cases for better automated program repair}. In \bibinfo{booktitle}{\emph{Proceedings of the 2017 11th joint meeting on foundations of software engineering}}. \bibinfo{pages}{831--841}.
\newblock


\bibitem[Yuan et~al\mbox{.}(2024)]%
        {yuan2024manual}
\bibfield{author}{\bibinfo{person}{Zhiqiang Yuan}, \bibinfo{person}{Yiling Lou}, \bibinfo{person}{Mingwei Liu}, \bibinfo{person}{Shiji Ding}, \bibinfo{person}{Kaixin Wang}, \bibinfo{person}{Yixuan Chen}, {and} \bibinfo{person}{Xin Peng}.} \bibinfo{year}{2024}\natexlab{}.
\newblock \bibinfo{title}{No More Manual Tests? Evaluating and Improving ChatGPT for Unit Test Generation}.
\newblock
\newblock
\showeprint[arxiv]{2305.04207}~[cs.SE]


\bibitem[Zan et~al\mbox{.}(2023)]%
        {zan2023large}
\bibfield{author}{\bibinfo{person}{Daoguang Zan}, \bibinfo{person}{Bei Chen}, \bibinfo{person}{Fengji Zhang}, \bibinfo{person}{Dianjie Lu}, \bibinfo{person}{Bingchao Wu}, \bibinfo{person}{Bei Guan}, \bibinfo{person}{Wang Yongji}, {and} \bibinfo{person}{Jian-Guang Lou}.} \bibinfo{year}{2023}\natexlab{}.
\newblock \showarticletitle{Large Language Models Meet {NL}2{C}ode: A Survey}. In \bibinfo{booktitle}{\emph{Proceedings of the 61st Annual Meeting of the Association for Computational Linguistics (Volume 1: Long Papers)}}, \bibfield{editor}{\bibinfo{person}{Anna Rogers}, \bibinfo{person}{Jordan Boyd-Graber}, {and} \bibinfo{person}{Naoaki Okazaki}} (Eds.). \bibinfo{publisher}{Association for Computational Linguistics}, \bibinfo{address}{Toronto, Canada}, \bibinfo{pages}{7443--7464}.
\newblock
\urldef\tempurl%
\url{https://doi.org/10.18653/v1/2023.acl-long.411}
\showDOI{\tempurl}


\bibitem[Zhang et~al\mbox{.}(2023)]%
        {pmlr-v202-zhang23av}
\bibfield{author}{\bibinfo{person}{Tianyi Zhang}, \bibinfo{person}{Tao Yu}, \bibinfo{person}{Tatsunori Hashimoto}, \bibinfo{person}{Mike Lewis}, \bibinfo{person}{Wen-Tau Yih}, \bibinfo{person}{Daniel Fried}, {and} \bibinfo{person}{Sida Wang}.} \bibinfo{year}{2023}\natexlab{}.
\newblock \showarticletitle{Coder Reviewer Reranking for Code Generation}. In \bibinfo{booktitle}{\emph{Proceedings of the 40th International Conference on Machine Learning}} \emph{(\bibinfo{series}{Proceedings of Machine Learning Research}, Vol.~\bibinfo{volume}{202})}, \bibfield{editor}{\bibinfo{person}{Andreas Krause}, \bibinfo{person}{Emma Brunskill}, \bibinfo{person}{Kyunghyun Cho}, \bibinfo{person}{Barbara Engelhardt}, \bibinfo{person}{Sivan Sabato}, {and} \bibinfo{person}{Jonathan Scarlett}} (Eds.). \bibinfo{publisher}{PMLR}, \bibinfo{pages}{41832--41846}.
\newblock
\urldef\tempurl%
\url{https://proceedings.mlr.press/v202/zhang23av.html}
\showURL{%
\tempurl}


\end{thebibliography}



\end{document}